\documentclass{optica-article}

\journal{opticajournal} 

\articletype{Research Article}
\usepackage[section]{placeins}
\usepackage{graphicx}     
\usepackage{subcaption}   
\usepackage{booktabs}
\usepackage{tabularx}
\usepackage{makecell}
\usepackage{placeins}
\usepackage{fancyhdr}
\providecommand{\JournalTitle}[1]{#1}

\definecolor{omittext}{RGB}{255,0,0}  
\usepackage[normalem]{ulem}
\newcommand{\newtext}[1]{\textcolor{omittext}{#1}} 
\usepackage{xcolor}

\usepackage{hyperref}
\usepackage{xcolor} 

\AtBeginDocument{%
  \hypersetup{
    colorlinks=true,
    linkcolor=blue,      
    citecolor=blue,      
    urlcolor=blue        
  }%
}

\begin{document}

\title{Hybrid Si-GST Polarization-Insensitive Dynamically Tunable Bifocal Metalens Operating at 1.55-$\mu$m Wavelength}

\author{Dipika Rani Nath,\authormark{1,2} Sadid Muneer\authormark{3} and Sajid Muhaimin Choudhury,\authormark{1,*}}

\address{\authormark{1}Department of Electrical and Electronic Engineering, Bangladesh University of Engineering and
Technology, Dhaka 1205, Bangladesh\\
\authormark{2}Department of Computer Science and Engineering, BRAC University, Dhaka 1212, Bangladesh\\
\authormark{3}Department of Electrical and Electronic Engineering, United International University, Dhaka 1212, Bangladesh}
\email{\authormark{*}sajid@eee.buet.ac.bd} 


\begin{abstract*} 
 
Metasurfaces have become a cornerstone of flat-optics, enabling precise control over light propagation through nanoengineered materials. Dynamic and reconfigurable metalenses are key to next-generation flat-optics platforms, yet their practical realization remains limited by slow response, optical loss, and polarization sensitivity. The integration of chalcogenide phase-change materials with metasurface architectures offers a powerful platform for dynamic optical tunability, owing to materials such as Ge$_2$Sb$_2$Te$_5$ (GST) that can be reversibly switched between amorphous and crystalline states with distinct refractive indices. However, the strong optical absorption of crystalline GST in the visible to near-infrared range has hindered its widespread use in reconfigurable metalenses. In this study, we design an all-dielectric polarization-insensitive metasurface based on hybrid Si-GST nanostructures to realize a dynamically tunable bifocal metalens operating at 1.55 \textmu m. The device achieves a variable focal length from 70 \textmu m to 200 \textmu m, with focusing efficiencies of 30\% in the amorphous state and 20\% in the crystalline state, as validated through finite-difference time-domain (FDTD) simulations. Using COMSOL Multiphysics, we show that flat-top laser excitation enables uniform, reversible phase transitions within tens of nanoseconds, amorphization in ~8 ns and crystallization in ~90 ns, without mechanical motion or electrical bias. For next-generation metasurfaces intended for uses including beam steering, dynamic holography, optical routing, multi-depth imaging, and optical communication, this method shows great promise due to its control and stability.
\end{abstract*}

\section{Introduction}

The capability to precisely modulate the amplitude, phase, and polarization of electromagnetic (EM) waves at subwavelength scales has fundamentally transformed modern photonics~\cite{r1}. Metasurfaces, the two-dimensional analogs of bulk metamaterials, consist of engineered arrays of subwavelength elements, often referred to as meta-atoms that can locally tailor optical wavefronts with exceptional precision~\cite{yu2014flat,jahani2016all}. Their compactness, scalability, and flexible design have enabled a broad range of optical functionalities, including holographic imaging~\cite{li2016multicolor,zheng2015metasurface}, optical cloaking~\cite{r2,r3}, polarization manipulation~\cite{r4, r5}, and spin–orbit interaction–based beam steering~\cite{yin2013photonic, r6}. Metalenses, or flat lenses composed of arrays of nanostructures, are one of the numerous types of metasurface designs that have emerged as an effective replacement for conventional curved lenses~\cite{r7,r8}. Metalenses can focus and manipulate light with remarkable precision while significantly lowering thickness and weight by carefully managing the light's phase through the geometry and materials of each nanostructure ~\cite{r9,khorasaninejad2016metalenses,chen2018broadband}. They are therefore well-suited for next-generation technologies including integrated photonic systems, wearable optics, and on-chip imaging~\cite{r10}.

However, most metalenses currently in use are static; once manufactured, their optical characteristics are fixed, which prevents many contemporary applications from adopting adaptive focusing or real-time tuning~\cite{r11}. Researchers have explored several tuning techniques to address this limitation. Stretchable substrates can change the focus length but may exhibit fatigue over time~\cite{r13}. Electrical tuning is possible with liquid-crystal-based designs, but they need complicated alignment and large layers~\cite{r14}. Thermo-optic and electro-optic techniques that use graphene or semiconductors frequently require a lot of power or offer little modulation~\cite{r15}. On the other hand, MEMS-based or multi-layer diffractive devices increase fabrication complexity~\cite{genevet2015holographic}. These difficulties emphasize the necessity of a small, quick, energy-efficient tuning mechanism that operates only in the metasurface plane. In this regard, phase-change materials (PCMs), particularly chalcogenide alloys, offer an attractive pathway toward dynamically reconfigurable metasurfaces. Their reversible and nonvolatile transitions between amorphous and crystalline states provide large and controllable refractive-index contrast without mechanical motion~\cite{rios2015integrated}. Among the chalcogenides, Ge$_2$Sb$_2$Te$_5$ (GST) has been the most extensively studied due to its large optical contrast, ultrafast switching speed, and compatibility with CMOS processes~\cite{shportko2008resonant,cao2020tunable}. Nevertheless, the relatively high absorption in crystalline GST limits optical transmission and overall device efficiency, posing a major bottleneck for transmissive PCM-based metalenses~\cite{r16}.
Despite significant progress in phase-change metasurfaces, most existing GST-based metalenses remain limited by non-uniform thermal switching, polarization dependence, and low focusing efficiency. 
\\Here, we introduce a hybrid Si--Ge$_2$Sb$_2$Te$_5$ (GST) all-dielectric metalens designed for dynamic focus at the telecommunication wavelength $\lambda_0 = 1550~\mathrm{nm}$ that achieves polarization-insensitive, all-optical varifocal switching through a flat-top laser excitation scheme. 
Unlike previous thermally or electrically driven approaches, this design enables uniform and reversible GST phase transitions within nanoseconds, eliminating the need for mechanical or electronic modulation. 
Finite-difference time-domain (FDTD) and coupled electromagnetic--thermal simulations confirm tunable dual-focal operation at 70~\textmu m and 200~\textmu m with focusing efficiencies up to 30\%. Owing to the symmetric cross-section of the nanofins, the proposed metalens exhibits polarization-insensitive operation for both linearly and circularly polarized light. While most PCM-based switchable metalenses remain proof-of-concept and lack practical feasibility due to their reliance on thermal annealing ~\cite{r17} for phase transition, we employ flat-top laser pulses to actively induce reversible GST phase changes through controlled laser heating, enabling fast and practical tunability. The dynamic tunability is achieved through laser-induced phase transitions in the GST layer, where controlled optical heating drives reversible amorphous–crystalline transformations. By tailoring the spatial intensity distribution of the laser (Gaussian or flat-top), the local refractive index can be precisely modulated, enabling continuous adjustment of the focal length. This all-optical control scheme enables a low-loss, compact, and reconfigurable metalens, bridging the gap between static metasurface optics and adaptive photonic platforms. The proposed approach offers significant potential for multi-depth imaging ~\cite{r18,r19}, optical tomography technique~\cite{r20,r21},
optical data storage~\cite{r22,r23},  optical communications~\cite{r24}, and optical tweezers~\cite{r25}.

\section{Structural Design}

\begin{figure}[htbp]
    \centering
    \includegraphics[width=0.62\linewidth]{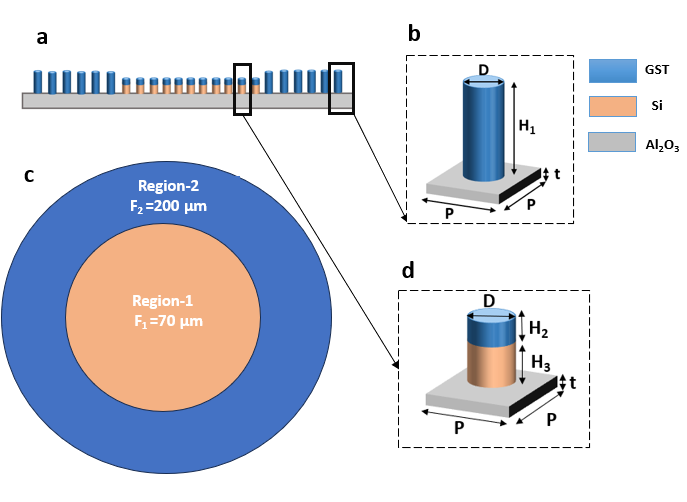}
    \caption{Schematic of the proposed varifocal metalens. 
(a) Top-view layout of the metalens illustrating the two concentric regions with focal lengths $f_1$ = 70 \textmu m and $f_2$ = 200 \textmu m. (b) Side view of the metasurface showing the spatial arrangement of nanopillars on an Al$_2$O$_3$ substrate.
(c) and (d) Enlarged views of the two nanopillar configurations used in Region~2 and Region~1, respectively, showing the geometric parameters, and different colors in the cells represent different materials.
 }
    \label{fig:img1}
\end{figure}

A metalens, a sophisticated optical device composed of arrays of nanoscale resonators that can individually adjust the amplitude and phase of transmitted light, is constructed from each meta-atom, or unit cell~\cite{kamali2018review}. By modifying its geometry, such as its diameter, height, or refractive-index contrast, each unit cell's phase shift may be precisely adjusted, giving accurate control over the way light passes through it. It is possible to form light waves in almost any way by arranging these meta-atoms in a particular phase pattern that is created from the desired optical profile. This allows for highly adaptable optical wavefronts through geometric phase engineering. The design of the proposed varifocal metalens is based on a dual-region configuration with two concentric zones engineered to exhibit distinct focal lengths under different material phase states, as illustrated in \textbf{Fig.~\ref{fig:img1}}a. The inner zone (Region 1) reaches up to a radius of 35~$\mu$m, whereas the outer region (Region 2) includes the region between 35~$\mu$m and 55~$\mu$m. In both regions, the transmitted light phase is precisely controlled by tightly arranged arrays of subwavelength nanopillars (meta-atoms). The proposed design integrates two distinct types of unit cells, as shown schematically in \textbf{Fig.~\ref{fig:img1}}b. The first type consists of a sapphire (Al$_{2}$O$_{3}$) substrate supporting a cylindrical Ge$_{2}$Sb$_{2}$Te$_{5}$ (GST) nanopillar, as illustrated in \textbf{Fig.~\ref{fig:img1}}c. The second type employs a hybrid GST-Si nanopillar configuration deposited on an Al$_{2}$O$_{3}$ substrate, as shown in \textbf{Fig.~\ref{fig:img1}}d.  This hybrid approach leverages the high refractive index of Si and the large, reversible index contrast of the GST to realize a dynamically reconfigurable metasurface. The cross-sectional schematics of the unit cells, shown in \textbf{Fig.~\ref{fig:img1}}(c,d), depict cylindrical nanopillars of diameter $D$ on an Al$_2$O$_3$ substrate.

All structural parameters are listed in \text{Table~\ref{tab:parameters}}. Here, $t$ denotes the thicknesses of the Al$_{2}$O$_{3}$ substrate, $H_{1}$, $H_{2}$ denote GST height in meta-atom \textbf{Fig.~\ref{fig:img1}}c and \textbf{Fig.~\ref{fig:img1}}d, respectively and  $H_{3}$ is Si height in meta-atom \textbf{Fig.~\ref{fig:img1}}d. $P$ denotes the period. 

\begin{table}[htbp]
\caption{Structural parameters
and their dimensions}
  \label{tab:parameters}
  \centering
\begin{tabular}{|c|c|}
\hline
Parameters & Dimension (nm)\\
\hline
$P$ & $600$ \\
$H_1$ & $1400$ \\
$H_2$ & $267$ \\
$H_3$ & $533$ \\
$t$ & $100$ \\
\hline
\end{tabular}
\end{table}
The optical constants of Si and Al$_{2}$O$_{3}$ were modeled using the Lumerical fit to Palik's data set~\cite{r26}, whereas the GST parameters were extracted from ellipsometric data at both amorphous and crystalline phases~\cite{r27}. At a wavelength of 1550~nm, the refractive index ($n$) and extinction coefficient ($k$) are approximately 2.4 and 0.02 for amorphous GST (a-GST), and 5.2 and 0.1 for crystalline GST (c-GST), respectively.

The propagation phase shift introduced by each nanopillar can be expressed as
\begin{equation}
\label{eq:eq1}
    \phi = \frac{2\pi n_{\mathrm{eff}} h}{\lambda},
\end{equation}
where $n_{\mathrm{eff}}$ represents the effective refractive index of the guided optical mode, $\lambda$ is the wavelength, and $h$ is the pillar height. By tuning the nanopillar diameter (D), $n_{\mathrm{eff}}$ can be continuously modulated to provide a complete $2\pi$ phase coverage. Due to their rotational symmetry, the designed nanopillars are polarization-insensitive under normal incidence, ensuring consistent focusing behavior for both linearly and circularly polarized light~\cite{r28,r29}.
\section{Methodology}
The dual-region configuration enables dynamic control of the focal position through all-optical tuning of the refractive index of the chalcogenide phase-change material Ge$_{2}$Sb$_{2}$Te$_{5}$ (GST), without requiring any mechanical movement or complex external control circuitry. In both regions, the transmitted light phase is precisely controlled by tightly arranged arrays of subwavelength nanopillars (meta-atoms). In this configuration, hybrid GST-silicon nanopillars on a sapphire substrate are integrated into Region 1, while optimized GST-only nanopillars constitute Region 2. Depending on the GST phase, the two region yields different optical responses. In contrast to the amorphous phase, GST's refractive index rises noticeably when it changes into its crystalline form. Changing the GST state alters the optical path length in each nanopillar, enabling a single metasurface to realize multiple phase profiles. The lens may dynamically change its focal length in response to thermal or optical stimuli by manipulating the spatial phase distribution across the surface for both configurations. The desired phase distribution for each region is derived from the hyperboloidal lens phase equation, which ensures constructive interference of the transmitted wavefronts at the designed focal point. For a target focal length $f$, the phase shift at a coordinate $(x,z)$ on the metasurface is expressed as:
\begin{equation}
\label{eq:phase_general}
\phi(x, z) = -\frac{2\pi}{\lambda} \left(\sqrt{x^2 + z^2 + f^2} - f\right),
\end{equation}
where $\lambda$ is the design wavelength (1550~nm in this case). This relation defines the phase retardation required to transform an incident planar wave into a converging spherical wavefront focused at distance $f$ from the metasurface. 

\begin{figure}[htbp]
    \centering
    \includegraphics[width=0.62\linewidth]{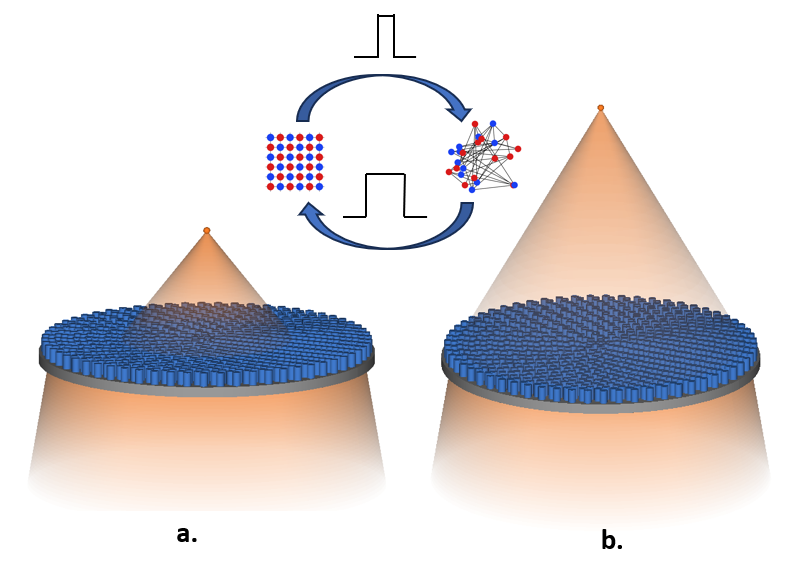}
    \caption{Schematic representation of the all-optical switching mechanism in the proposed varifocal metalens. 
(a) A longer rectangular laser pulse induces crystallization of the amorphous Ge$_2$Sb$_2$Te$_5$ (GST) layer by heating it above the glass transition temperature and enabling atomic rearrangement into the crystalline phase. 
(b) In contrast, a short and intense laser pulse rapidly melts and quenches the GST layer, converting it back to the amorphous state. 
These reversible phase transitions modulate the refractive index of the metasurface, enabling dynamic switching between two distinct focal states.}
    \label{fig:img6}
\end{figure}

For the inner region (Region~1), the phase profile in Eq.~\ref{eq:phase_general} is used with $f_{1}=70~\mu$m. The hybrid GST-Si nanopillars are geometrically optimized to produce the required phase coverage in the crystalline state, where the higher refractive index of GST enables a complete $2\pi$ phase modulation within a single layer. The diameter of each nanopillar is varied to adjust the effective refractive index $n_{\text{eff}}$ of the guided optical mode, which determines the local phase delay according to Eq.~\ref{eq:eq1}. When GST is in the crystalline state, this inner region focuses incident light tightly at a focal distance of 70~$\mu$m. The outer region (Region~2) follows a similar design principle but is optimized for a longer focal length $f_{2}=200~\mu$m. Its phase profile is defined by Eq.~\ref{eq:phase_general} with $f = f_{2}$, and the unit cells consist entirely of GST nanopillars fabricated directly on the Al$_{2}$O$_{3}$ substrate. When the material is in the amorphous state, the lower refractive index of GST shifts the optical path difference, allowing Region~2 to satisfy the phase condition for focusing at $f_{2}=200~\mu$m, while the crystalline-state Region~1 becomes inactive. When GST is crystalline, only Region~1 (hybrid GST–Si) satisfies the phase-matching condition, yielding a short focal length of $f_{1}=70~\mu$m. This complementary configuration ensures that the lens exhibits dual-focal functionality as shown in \textbf{Fig.~\ref{fig:img6}}. Thus, the same metasurface can dynamically toggle between two distinct focusing states by leveraging the nonvolatile phase transition behavior of GST. This design principle forms the foundation of a compact, efficient, and purely optical varifocal system, capable of fast switching between focal planes for use in adaptive imaging, LiDAR, and reconfigurable photonics integration~\cite{qian2020broadband,park2025end,wang2015quasi}.

\section{Result}\label{sec:Method}

This section encompasses three primary analyses. First, the transmittance and phase responses of the unit cells were examined using finite-difference time-domain (FDTD) simulations. Second, the bifocal performance of the complete metalens was analyzed in both amorphous and crystalline states. Finally, an opto-thermal study was carried out using COMSOL Multiphysics to investigate the laser-induced phase switching behavior of the GST layer.

\subsection{Optical response of the unit cell}

\begin{figure}[htbp]
    \centering
    \begin{subfigure}[b]{0.43\linewidth}
        \centering
        \textbf{(a)}\\[0pt] 
        \includegraphics[width=\linewidth]{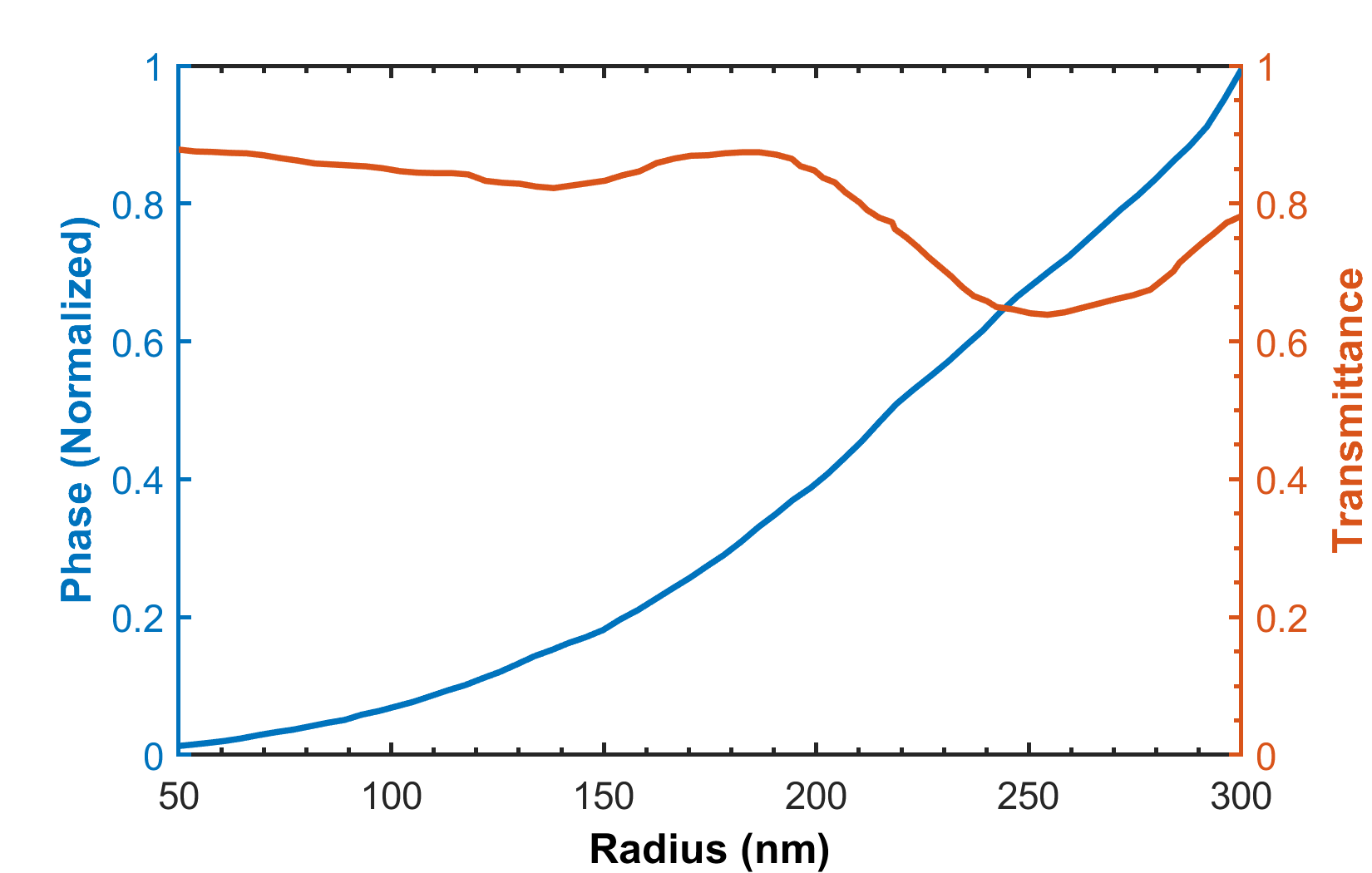}
        
        \label{fig:img4}
    \end{subfigure}
    \hspace{2mm}
    \begin{subfigure}[b]{0.43\linewidth}
        \centering
        \textbf{(b)}\\[0pt] 
        \includegraphics[width=\linewidth]{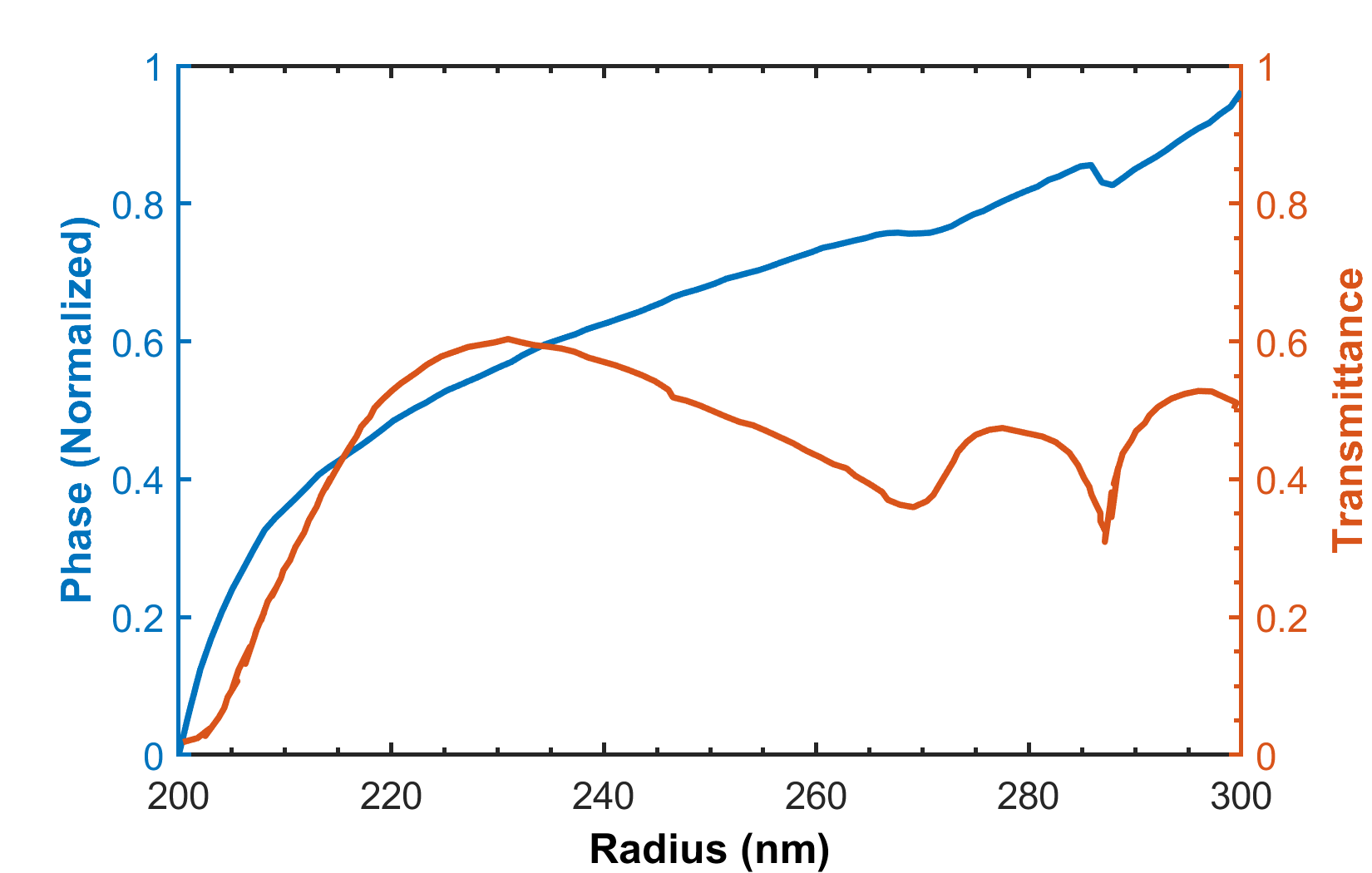}

        \label{fig:img5}
    \end{subfigure}

    \caption{Simulated transmittance and phase response of the metasurface unit cells as a function of nanopillar radius, obtained from three-dimensional FDTD simulations at a wavelength of 1550~nm. 
(a) Amorphous-state GST nanopillars with a height of 1400~nm exhibit complete $0{-}2\pi$ phase coverage and a transmission efficiency of approximately 85\%. 
(b) Crystalline-state GST nanopillars with a height of 800~nm also achieve full-phase modulation, though with slightly reduced transmittance due to the higher optical absorption of crystalline GST.}
    \label{fig:comparison}
\end{figure}

To realize an efficient focusing metasurface, each unit cell transmission phase must span a complete \(2\pi\) range. This is achieved by tailoring the effective refractive index through variations in the nanopillar diameter. Three-dimensional FDTD simulations were carried out using Lumerical FDTD Solutions, with periodic boundary conditions along the $x$ and $y$ directions and perfectly matched layers (PML) along $z$. An $x$-polarized plane wave at 1550~nm wavelength, propagating in the $+z$ direction, was used as the excitation source. A far-field monitor was employed to extract the transmitted intensity and phase distribution. For the amorphous state of GST, a pillar height of 1400~nm in \textbf{Fig.~\ref{fig:img1}}c yielded a full $0$–$2\pi$ phase coverage with a transmission efficiency of approximately 85\%, and by sweeping the pillar radius, the relation between radius and phase shift was extracted, as depicted in \textbf{Fig.~\ref{fig:comparison}}a. For the crystalline state, the optimized pillar height was 800~nm in \textbf{Fig.~\ref{fig:img1}}d, which also achieved full phase modulation but with reduced transmission due to the higher extinction coefficient of crystalline GST, as shown in \textbf{Fig.~\ref{fig:comparison}}b. The introduction of a Si base in the hybrid structure significantly improved transmittance compared to a pure GST pillar. The hybrid unit cell modulates the phase to achieve a focus at $f_{1}$ = 70 \textmu m in the crystalline state, whereas the GST-only unit cell achieves a focus at $f_{2}$ = 200 \textmu m in the amorphous state. These functionalities arise from the refractive-index contrast between the two GST phases, as shown in \href{Sup_Doc.pdf}{Supplement 1}. 

\subsection{Analysis of bifocal metalens}

\begin{figure}[htbp]
    \centering

    \begin{subfigure}[b]{0.4\linewidth}
        \centering
        \includegraphics[width=\linewidth]{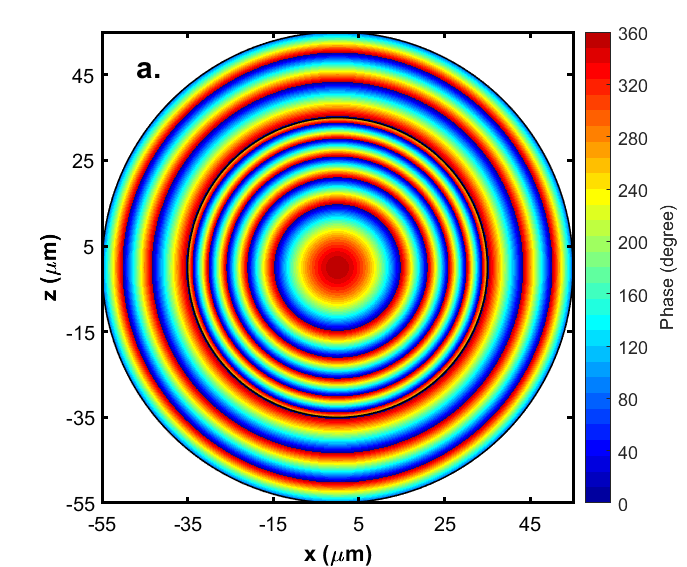}
    
        \label{fig:img21}
    \end{subfigure}
    \hspace{2mm}
    \begin{subfigure}[b]{0.40\linewidth}
        \centering
        \includegraphics[width=\linewidth]{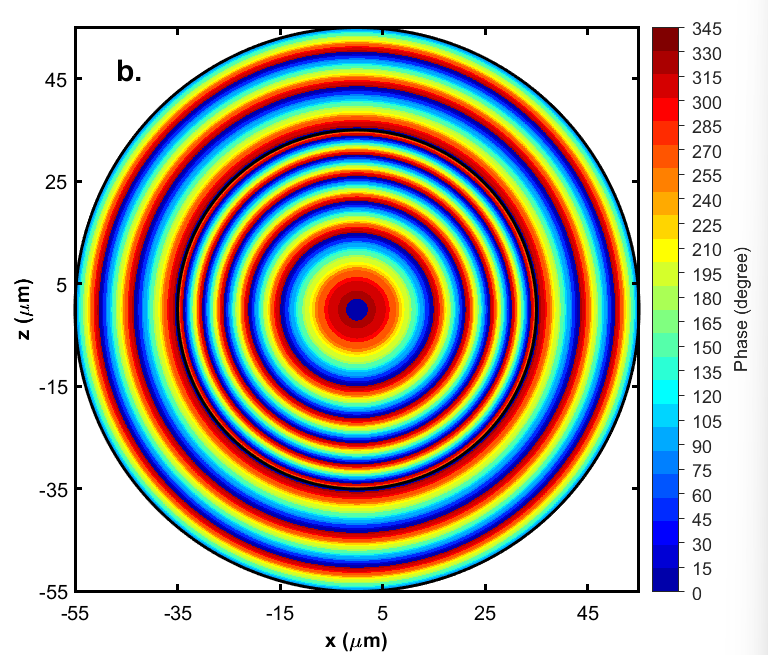}
    
        \label{fig:img22}
    \end{subfigure}

    \vspace{2mm} 

    \begin{subfigure}[b]{0.4\linewidth}
        \centering
        \includegraphics[width=\linewidth]{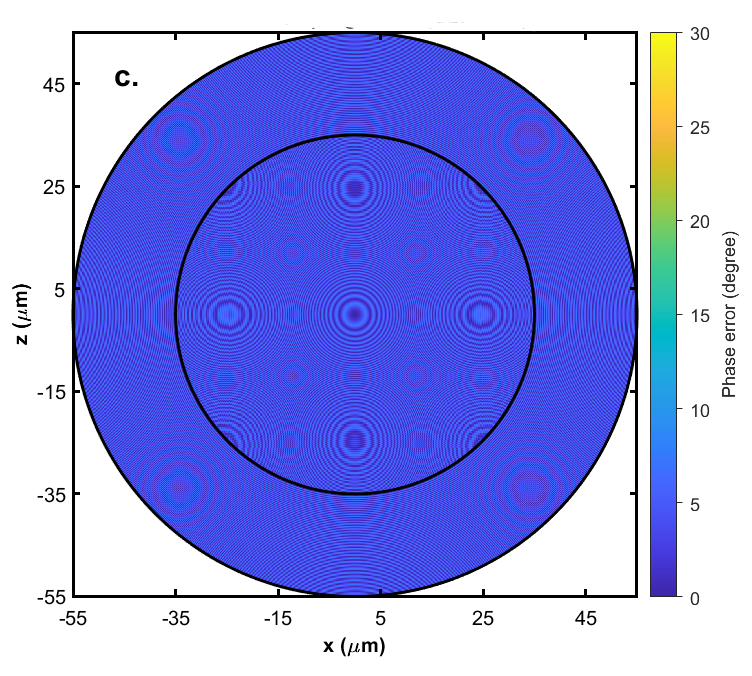}
   
        \label{fig:img23}
    \end{subfigure}
    \hspace{1mm}
    \begin{subfigure}[b]{0.37\linewidth}
        \centering
        \includegraphics[width=\linewidth]{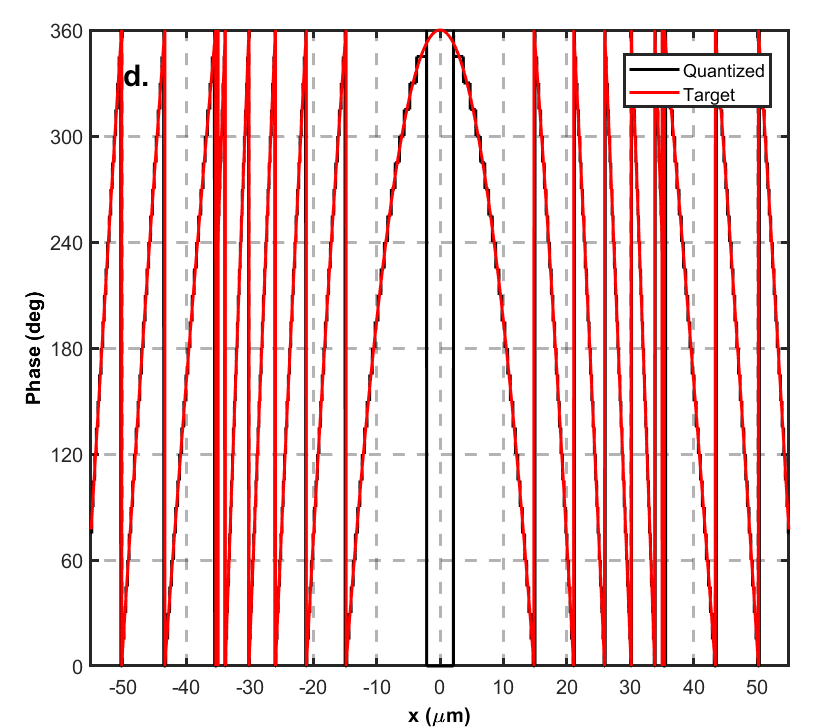}
      
        \label{fig:img24}
    \end{subfigure}

    \caption{ Numerical results of the dual-focal metasurface lens. 
  (a) Target continuous phase profile for the designed dual-focal configuration ($f_1 = 70~\mu$m, $f_2 = 200~\mu$m). 
  (b) Realized 24-level discretized phase distribution based on the nanopillar radii. 
  (c) Phase error map between the target and discretized profiles.
  (d) Comparison of the target (red) and quantized (black) phase variations along the central axis ($y = 0$).}
    \label{fig:phase}
\end{figure}

\begin{figure}[htbp]
    \centering

    \begin{subfigure}[b]{0.4\linewidth}
        \centering
        \includegraphics[width=\linewidth]{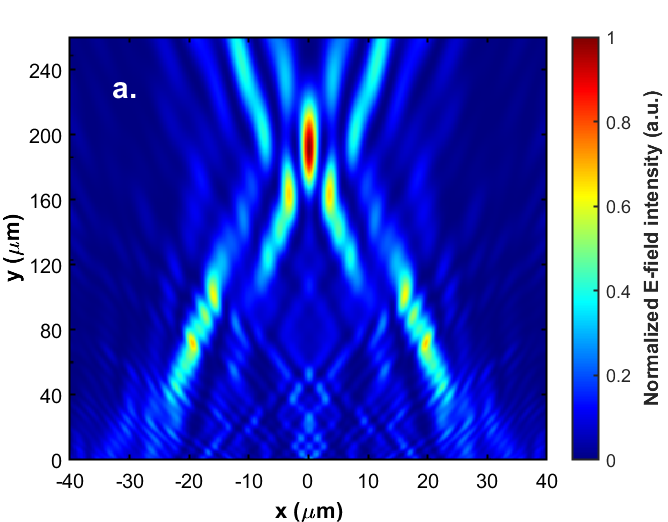}
    
        \label{fig:img15}
    \end{subfigure}
    \hspace{2mm}
    \begin{subfigure}[b]{0.4\linewidth}
        \centering
        \includegraphics[width=\linewidth]{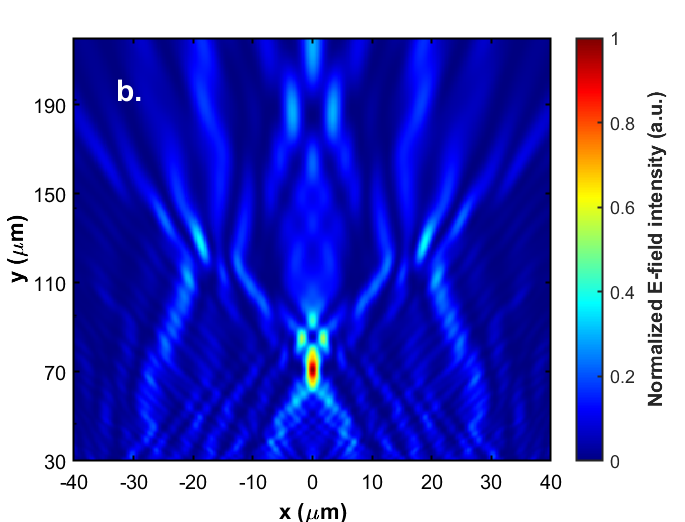}
    
        \label{fig:img16}
    \end{subfigure}

    \vspace{2mm} 

    \begin{subfigure}[b]{0.38\linewidth}
        \centering
        \includegraphics[width=\linewidth]{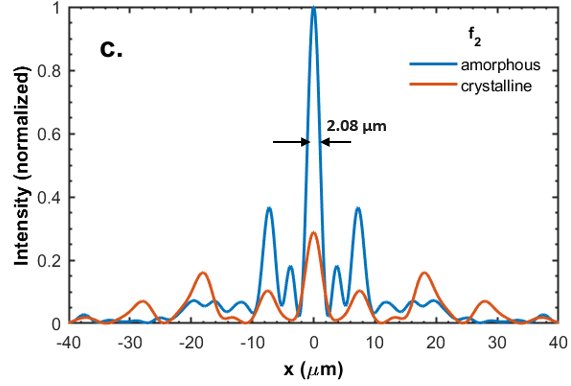}
   
        \label{fig:img14}
    \end{subfigure}
    \hspace{2mm}
    \begin{subfigure}[b]{0.38\linewidth}
        \centering
        \includegraphics[width=\linewidth]{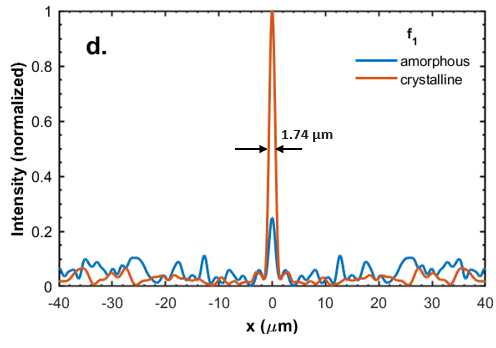}
      
        \label{fig:img13}
    \end{subfigure}

    \caption{Normalized electric-field intensity profiles along the $x{-}y$ plane in (a) amorphous and (b) crystalline states. The focal positions correspond to $f_2$ = 200 \textmu m and $f_1$ = 70 \textmu m, respectively, demonstrating the tunable focusing behavior of the metalens. 
(c) and (d) show the normalized intensity distributions of the focused beams along the $x$-axis for amorphous and crystalline configurations, respectively.}
    \label{fig:intensity}
\end{figure}

\begin{figure*}[htbp]
    \centering
    \begin{subfigure}[b]{0.23\textwidth}
        \includegraphics[width=\linewidth]{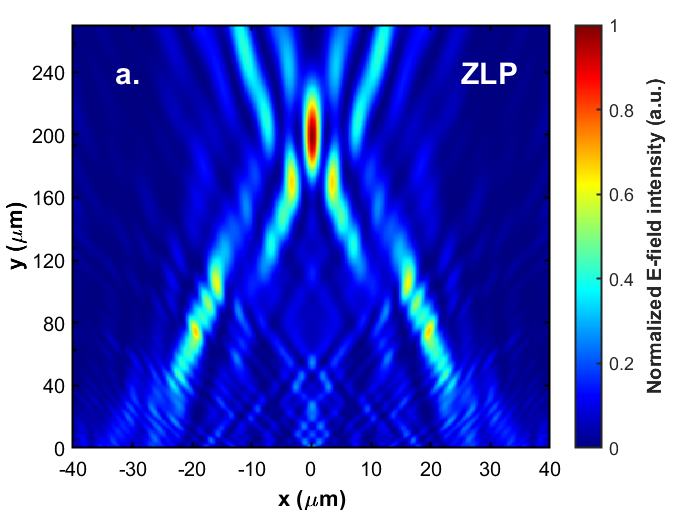}
       
    \end{subfigure}
    \hspace{0.1mm}
    \begin{subfigure}[b]{0.23\textwidth}
        \includegraphics[width=\linewidth]{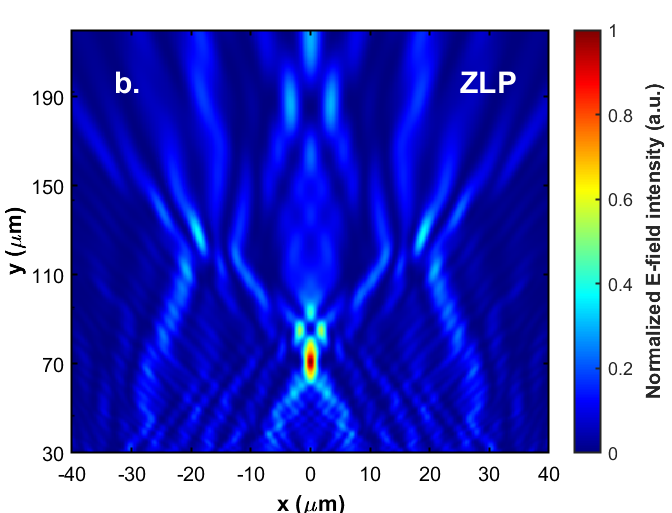}
       
    \end{subfigure}
    \hspace{0.1mm}
    \begin{subfigure}[b]{0.23\textwidth}
        \includegraphics[width=\linewidth]{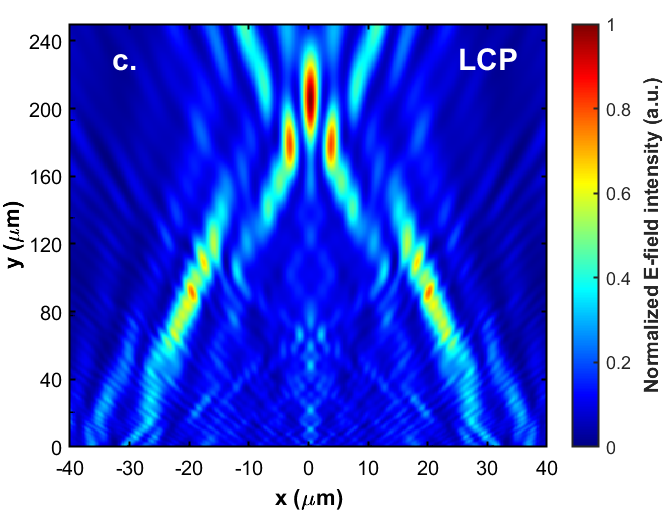}
       
    \end{subfigure}
    \hspace{0.1mm}
    \begin{subfigure}[b]{0.23\textwidth}
        \includegraphics[width=\linewidth]{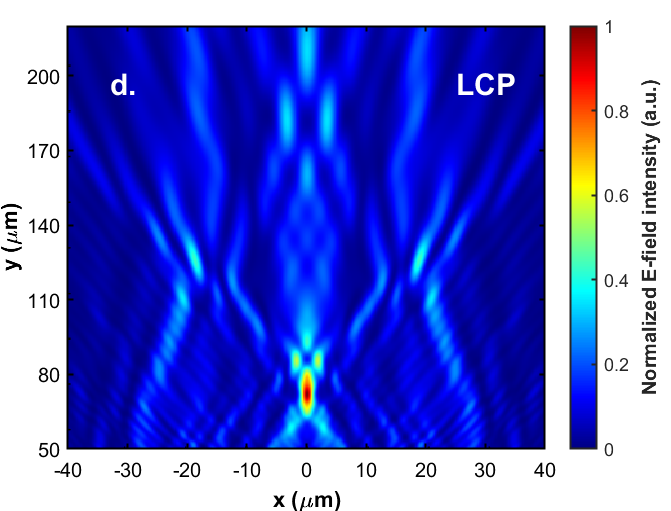}
  
    \end{subfigure}

    \vspace{1mm}
    \begin{subfigure}[b]{0.23\textwidth}
        \includegraphics[width=\linewidth]{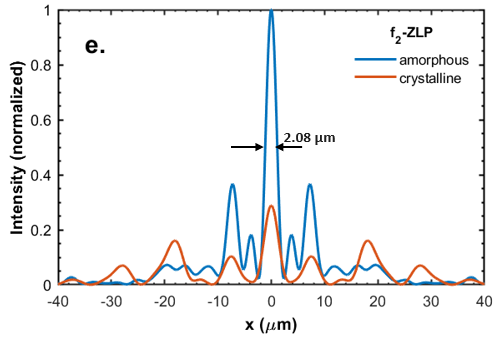}
     
    \end{subfigure}
    \hspace{0.1mm}
    \begin{subfigure}[b]{0.23\textwidth}
        \includegraphics[width=\linewidth]{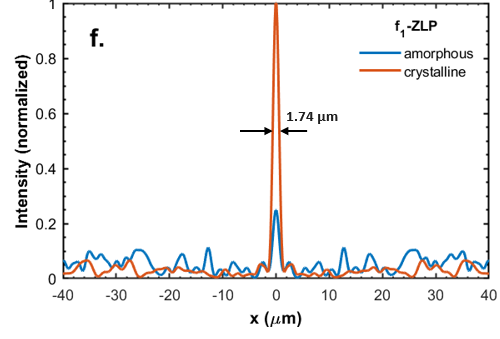}
    
    \end{subfigure}
    \hspace{0.1mm}
    \begin{subfigure}[b]{0.23\textwidth}
        \includegraphics[width=\linewidth]{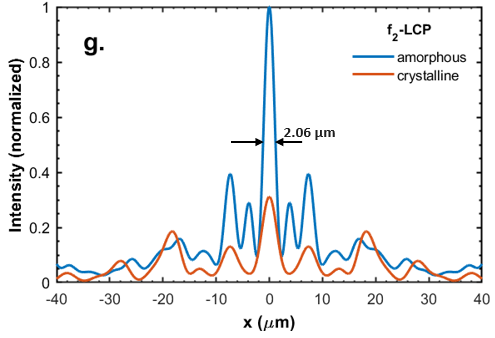}
       
    \end{subfigure}
    \hspace{0.1mm}
    \begin{subfigure}[b]{0.23\textwidth}
        \includegraphics[width=\linewidth]{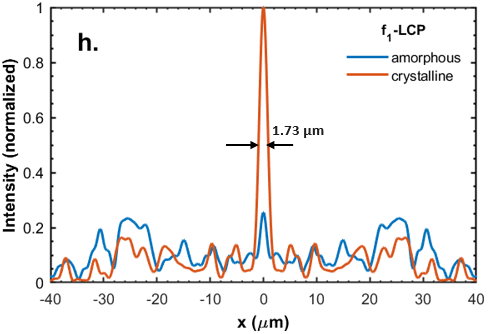}
     
    \end{subfigure}

\caption{
Simulated focusing performance of the proposed metalens under different polarization states. 
(a,~b) Normalized electric-field intensity distributions for ZLP illumination in amorphous and crystalline states. 
(c,~d) Corresponding results for LCP illumination. 
(e--h) Intensity profiles along the focal plane showing full width at half maximum (FWHM) values of 2.08~µm and 1.73~µm for amorphous and crystalline ZLP light, and 2.06~µm and 1.73~µm for amorphous and crystalline LCP light, respectively. 
The nearly identical focal responses confirm the polarization-insensitive behavior of the metalens.
}

    \label{fig:pol}
\end{figure*}

A dual-focal metalens was designed to adjust its focal length between 70~\textmu m and 200~\textmu m depending on the phase state of GST. \textbf{Fig.~\ref{fig:phase}} illustrates the simulated phase evolution and corresponding quantization effects for the proposed dual-focal metasurface lens, where the inner region ($f_1 = 70~\mu$m) and outer annular region ($f_2 = 200~\mu$m) jointly contribute to achieving dual-depth focusing. 
The target continuous phase distribution shown in \textbf{Fig.~\ref{fig:phase}}a demonstrates a smooth radial progression of the optical phase from the center to the periphery, corresponding to the analytical design that compensates for the propagation delay of both focal zones. 
By discretizing this continuous profile into 24 uniform phase levels (15$^{\circ}$ spacing) and mapping each level to its corresponding nanopillar radius, as listed in Tables S2 and S3 of \href{Sup_Doc.pdf}{Supplement 1}, the realized quantized phase pattern in \textbf{Fig.~\ref{fig:phase}}b  reproduces the analytical phase while introducing stepwise boundaries between adjacent phase levels. 
The phase error map presented in \textbf{Fig.~\ref{fig:phase}}c quantifies the deviation between the analytical and discretized phase values, revealing a maximum error below 30$^{\circ}$ within the entire aperture, which indicates high phase fidelity and minimal quantization loss. 
The comparison of the target and discretized phase profiles along the central line ($y=0$) in \textbf{Fig.~\ref{fig:phase}}d further confirms that the realized phase accurately follows the theoretical saw-tooth progression, maintaining the expected 0-360$^{\circ}$ periodicity with only minor rounding effects introduced by discrete sampling. \noindent
The electric field intensity distributions presented in \textbf{Fig.~\ref{fig:intensity}} and\textbf{~\ref{fig:pol}} were obtained using a two-dimensional finite-difference time-domain (FDTD) simulation performed in \textit{Lumerical FDTD Solutions}. 
To capture the near-field response, a one-dimensional field monitor was positioned just above the nanopillar array. 
Perfectly matched layers (PMLs) were applied along all boundaries to suppress unwanted reflections. 
An $x$-polarized plane wave propagating along the positive $y$-direction was used as the excitation source, while the transmitted electromagnetic field recorded by the monitor was projected along the positive $x$-direction for further analysis. The simulated $x$–$y$ intensity profiles for the amorphous and crystalline states are shown in \textbf{Fig.~\ref{fig:intensity}}a and \textbf{Fig.~\ref{fig:intensity}}b, respectively. When GST is amorphous, Region~II of the metalens satisfies the phase distribution defined by Eq.~\eqref{eq:phase_general}, yielding a focal point at 204~\textmu m. Upon crystallization, Region~I satisfies the same phase condition, shifting the focal length to 74~\textmu m. The satisfaction of these phase distributions under amorphous and crystalline states is presented in the \href{Sup_Doc.pdf}{Supplement 1}. Thus, a single device dynamically tunes its focal distance from 74~\textmu m ($f_{1}$) to 204~\textmu m ($f_{2}$). Minor deviations between the simulated and target focal positions result from discrete phase sampling across the metasurface. The contrast between the two focusing states is further verified by the normalized one-dimensional  electric-field intensity profiles (\(I/I_0\)) and corresponding full width at half maximum (FWHM) values along the $x$-axis of the metalens, illustrated in \textbf{Fig.~\ref{fig:intensity}}c and \textbf{Fig.~\ref{fig:intensity}}d. The theoretical diffraction-limited FWHM, estimated using 
$\mathrm{FWHM} = \lambda/(2\,\mathrm{NA})$, 
is 1.73 \textmu m ($f_1$). 
The simulated focal spot exhibits a FWHM of 1.74 \textmu m, indicating near-diffraction-limited focusing for $f_1$. 
For $f_2$, considering the annular aperture (35--55\textmu m) 
and using the analytical relation 
$\mathrm{FWHM}_{\text{ring}} = 0.357\,(\lambda f / R)$~\cite{khonina2013sharper}, 
the estimated value is 2.09 \textmu m.
Thus, this
design validates near-diffraction-limited focusing, while the simulated focal spot shows a FWHM of $2.06~\mu\text{m}$.
 The calculated focusing efficiencies are 20\% ($f_1$) and 30\% ($f_2$), which is defined as the fraction of incident optical power being concentrated within a radius of 3 × FWHM on the focal plane around the focal
axis~\cite{shalaginov2021reconfigurable,zhuang2019high}. The slight reduction in focusing efficiency and modulation range of the varifocal metalens, compared with the single-focal design, can be attributed to the shared-aperture configuration. Upon phase transition, the modulation of focusing intensity $(\frac{\Delta I}{I_0}
)$ reaches 78.13\% ($f_1$) and 71.23\% ($f_2$). validating the varifocal capability of the proposed metalens~\cite{yu2026multifunctional}. \noindent
The polarization-insensitive performance of the designed metalens was investigated by analyzing its focusing characteristics under linearly and circularly polarized illumination. 
As illustrated in \textbf{Fig.~\ref{fig:pol}}(a--d), the metalens exhibits comparable focal behavior for Z-linearly polarized light (ZLP) and left circular polarized light (LCP) excitations in both the amorphous and crystalline states, confirming its polarization-independent response. 
The corresponding intensity profiles in \textbf{Fig.~\ref{fig:pol}}(e--h) reveal full width at half maximum (FWHM) values of 2.06~µm and 1.74~µm for amorphous and crystalline ZLP light, respectively. 
Similarly, for LCP illumination, the measured FWHM values are 2.08~µm in the amorphous state and 1.75~µm in the crystalline state. 
The slightly narrower focal spot observed under circular polarization arises from its more symmetric field distribution\cite{richards1959electromagnetic}. 
These results verify that the metasurface maintains consistent and efficient focusing across both polarization states and phase-change conditions.

\newtext{To quantitatively evaluate the bifocal performance, as summarized in  \text{Table~\ref{tab:selectivity}}. the focal selectivity is defined as the ratio between the peak intensity at the desired focal plane and the residual intensity at the undesired focal plane. For the $f_1$ mode in the crystalline state, the extracted selectivity is 20, indicating strong suppression of the inactive focal channel. For the $f_2$ mode in the amorphous state, the corresponding selectivity is 3.3, indicating that the desired long-focus mode remains dominant despite a residual contribution from the inactive state. In addition, the focal contrast, defined as the ratio between the focal peak intensity and the surrounding background intensity, is estimated to lie in the range of 20--50, confirming well-confined focusing behavior. These results demonstrate effective bifocal switching in the proposed metalens, with stronger focal discrimination observed for the $f_1$ mode.}

\begin{table}[t]
\centering
\caption{\textcolor{red}{Quantitative evaluation of bifocal selectivity for the proposed metalens.}}
\label{tab:selectivity}
\color{red}
\begin{tabular}{|c|c|c|c|}
\hline
\textbf{State} & $\mathbf{I(f_1)}$ & $\mathbf{I(f_2)}$ & \textbf{Selectivity} \\
\hline
Crystalline & 1   & 0.05 & 20 \\
\hline
Amorphous   & 0.3 & 1    & 3.3 \\
\hline
\end{tabular}
\end{table}


\begin{figure}[t]
    \centering
    \includegraphics[width=0.6\textwidth]{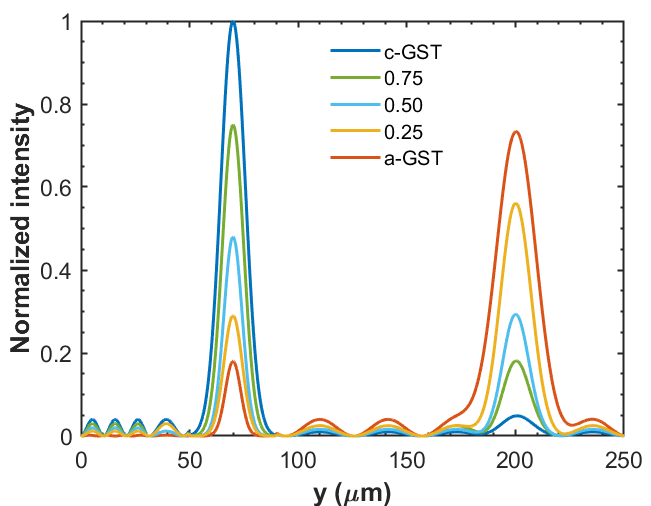}
 \caption{\newtext{Axial intensity profiles of the bifocal metalens for crystalline (c-GST), amorphous (a-GST), and intermediate states (25\%, 50\%, and 75\% crystallization). Two focal positions at $f_1 \approx 70\,\mu$m and $f_2 \approx 200\,\mu$m are observed, with a gradual shift in dominance between them. The results highlight tunable bifocal behavior and reduced selectivity under partial crystallization.}}
    \label{fig:your_label}
\end{figure}

\begin{figure}[t]
    \centering

    \begin{subfigure}[b]{0.3\textwidth}
        \centering
        \includegraphics[height=4.5cm]{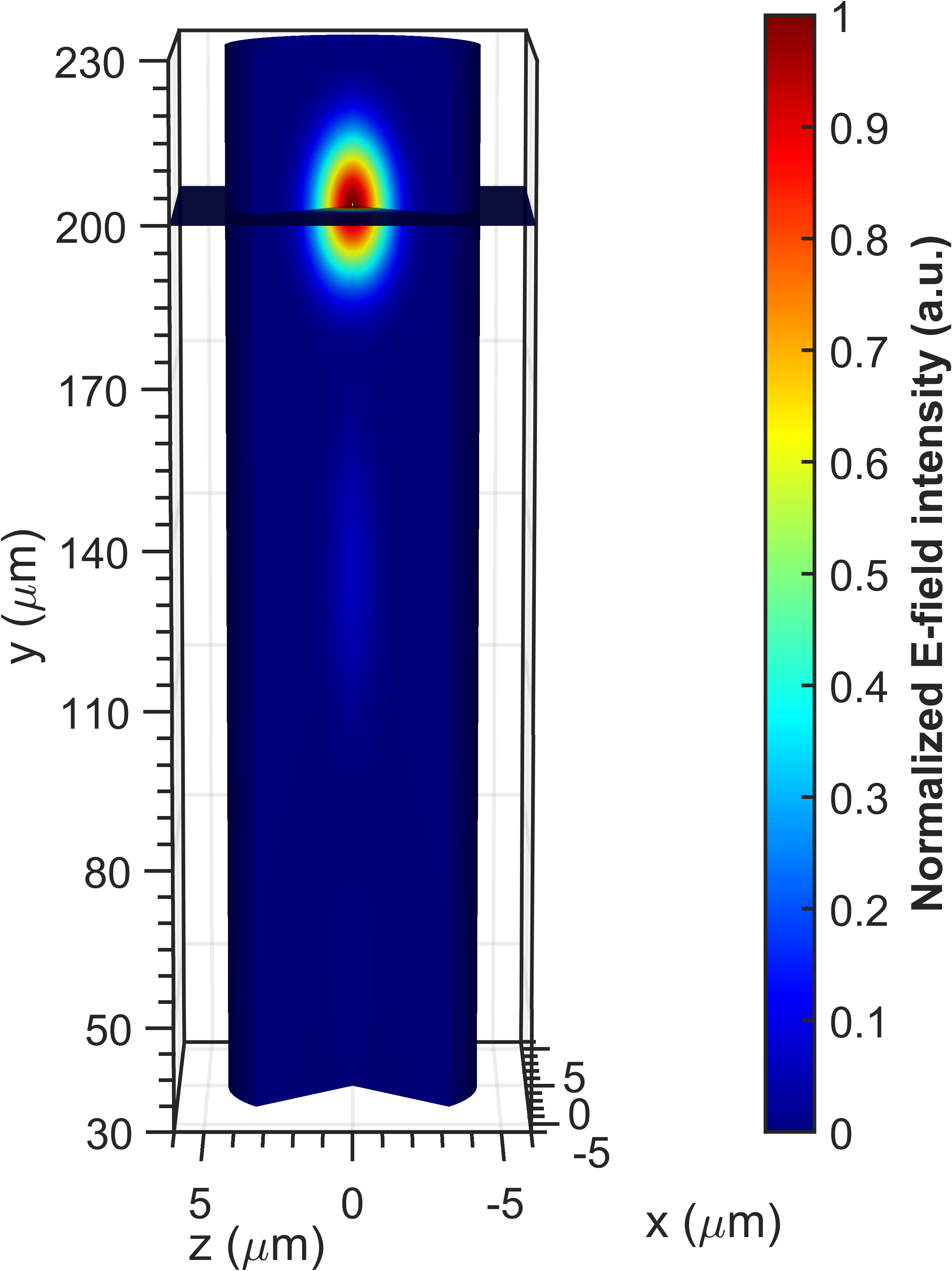}
        \caption{}
    \end{subfigure}
      \hspace{0.1mm}
    \begin{subfigure}[b]{0.3\textwidth}
        \centering
        \includegraphics[height=4.5cm]{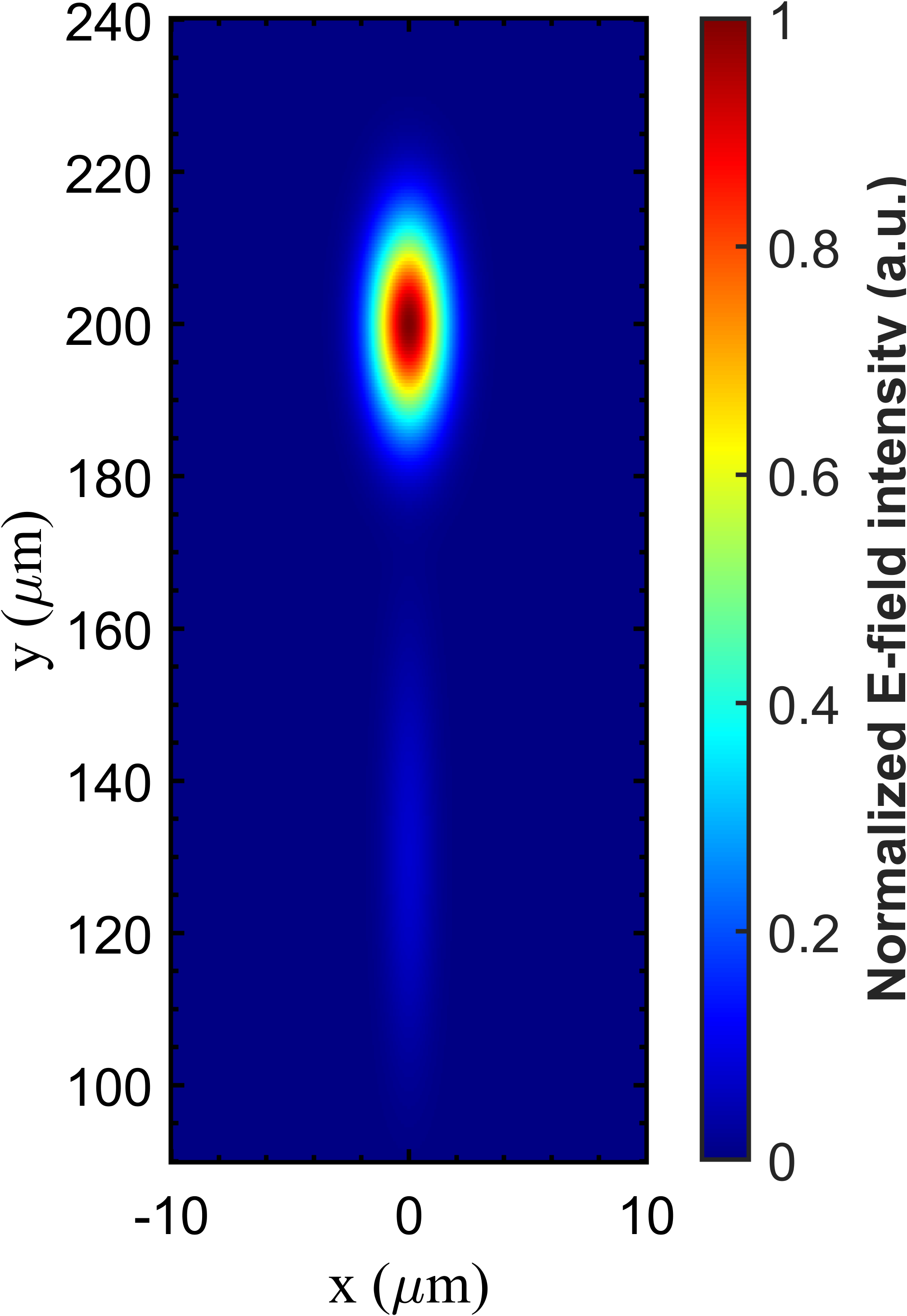}
        \caption{}
    \end{subfigure}
      \hspace{0.1mm}
    \begin{subfigure}[b]{0.3\textwidth}
        \centering
        \includegraphics[height=4.5cm]{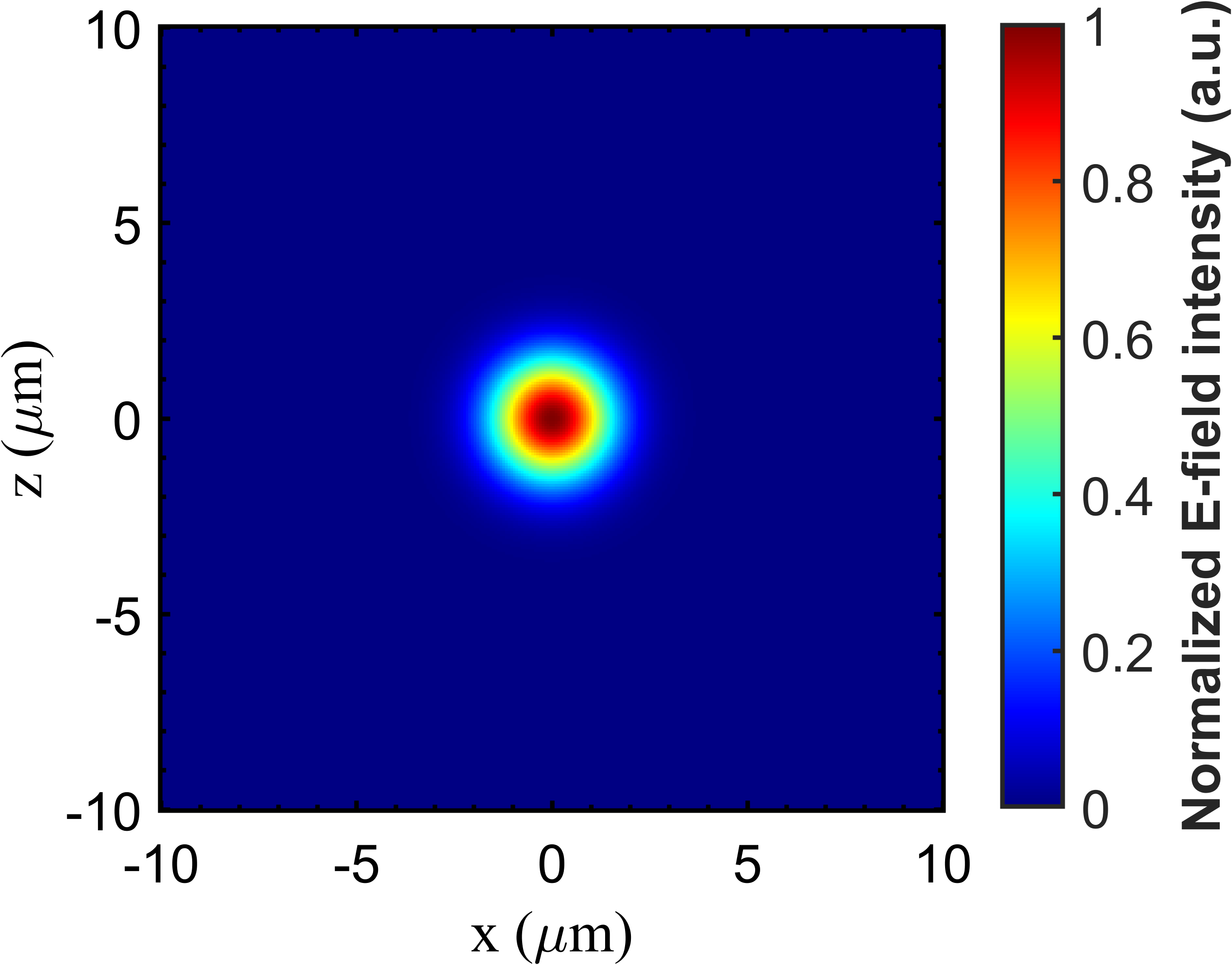}
        \caption{}
    \end{subfigure}

    \vspace{1mm}

    \begin{subfigure}[b]{0.3\textwidth}
        \centering
        \includegraphics[height=4.5cm]{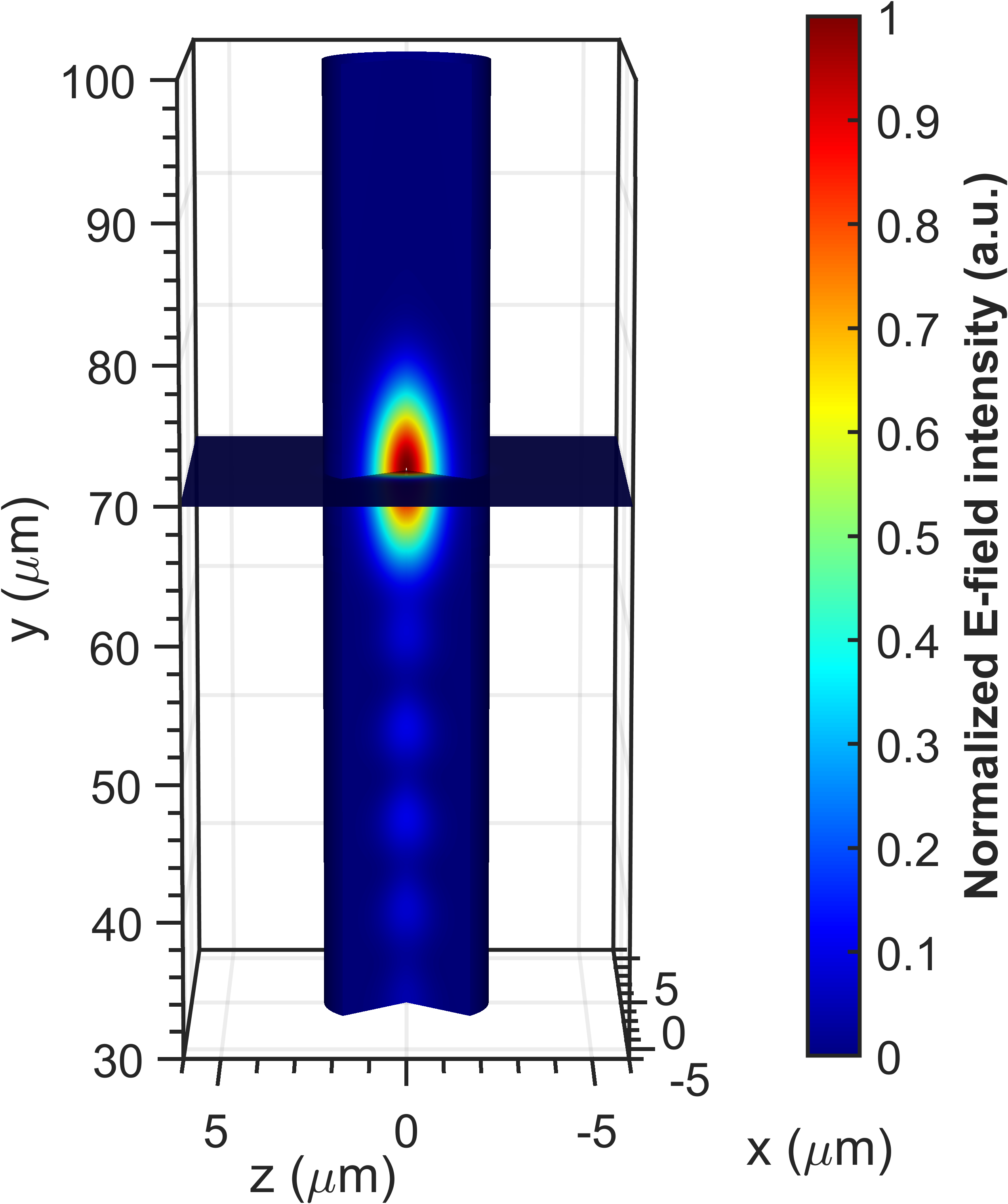}
        \caption{}
    \end{subfigure}
      \hspace{0.1mm}
    \begin{subfigure}[b]{0.3\textwidth}
        \centering
        \includegraphics[height=4.5cm]{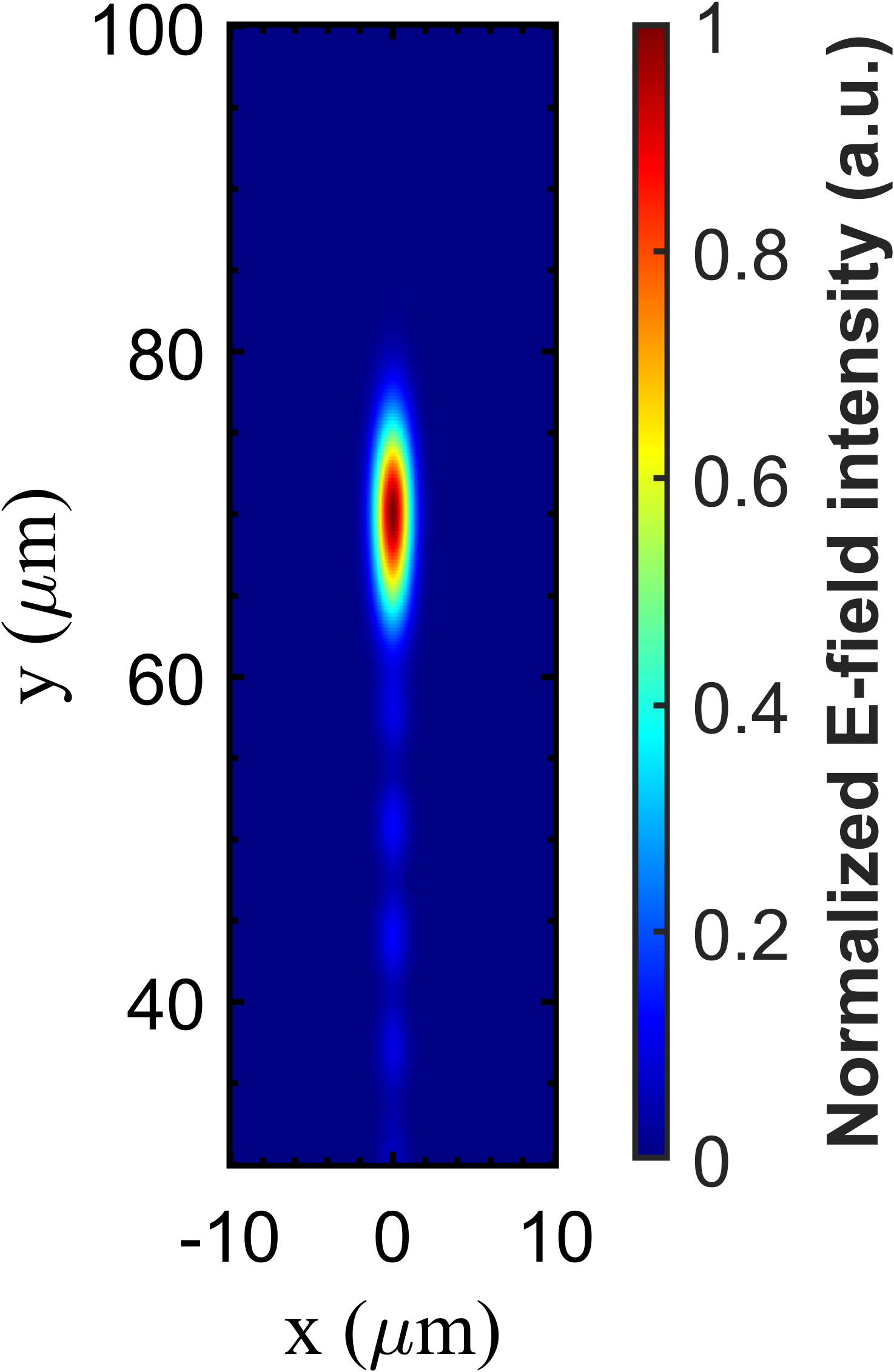}
        \caption{}
    \end{subfigure}
      \hspace{0.1mm}
    \begin{subfigure}[b]{0.3\textwidth}
        \centering
        \includegraphics[height=4.5cm]{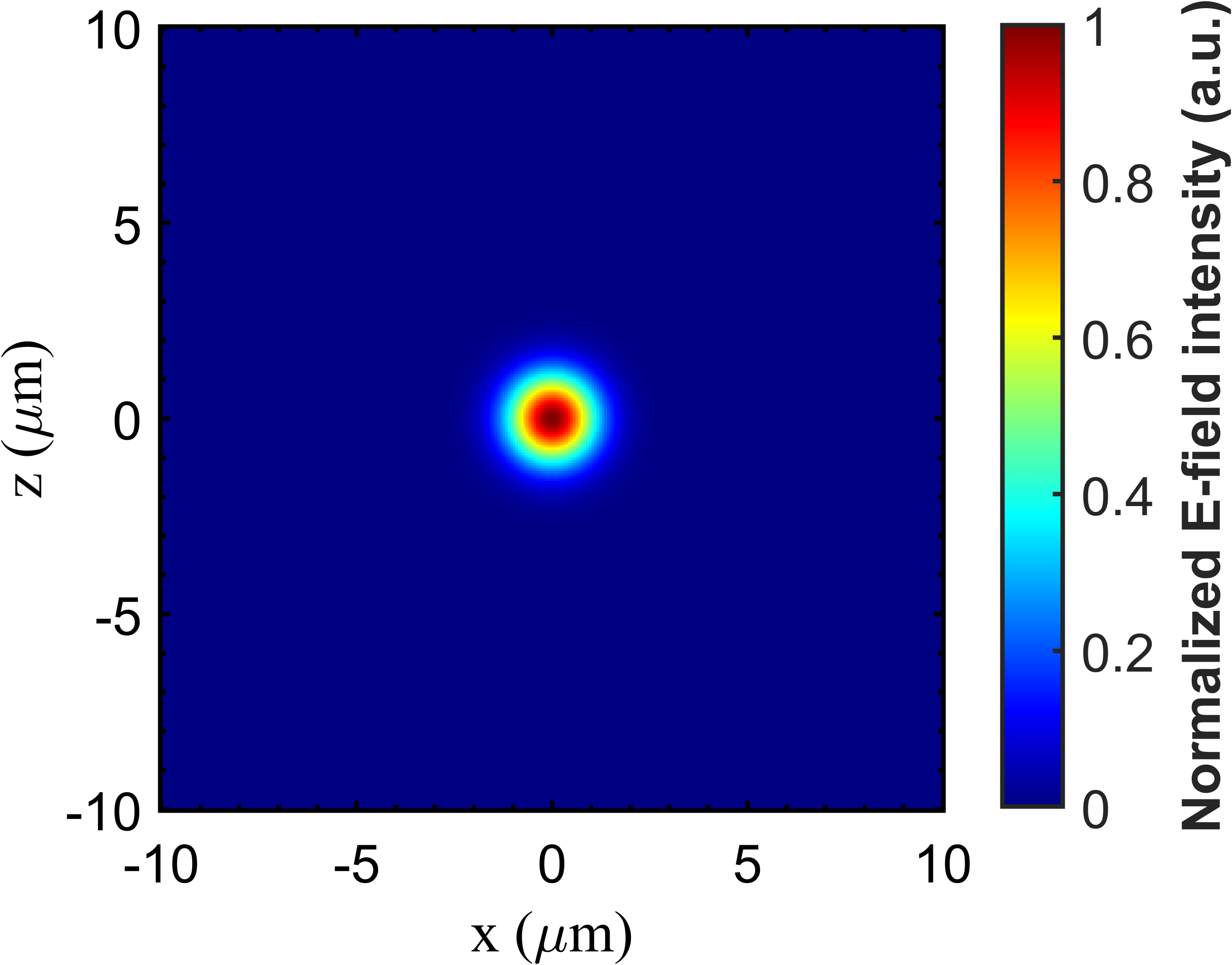}
        \caption{}
    \end{subfigure}

    \caption{\newtext{Full-lens 3D FDTD validation of the bifocal metalens in the two GST states. (a) Three-dimensional normalized electric-field intensity distribution in the amorphous state, showing focusing near $200\,\mu$m. (b) Corresponding axial $x$--$y$ intensity map. (c) Transverse intensity profile at the focal plane, confirming a localized focal spot. (d)--(f) Corresponding 3D field distribution, axial intensity map, and focal-plane intensity profile for the crystalline state, showing focusing near $70\,\mu$m.}}
    \label{fig:row_images}
\end{figure}
\newtext{To investigate the impact of partial phase transitions, we evaluated the focusing behavior for intermediate GST states with different crystallization fractions. As shown in \textbf{Fig.~\ref{fig:your_label}}, the focal intensity at the designed focal length decreases progressively as the GST transitions from the fully crystalline to the amorphous state. Simultaneously, residual intensity emerges at the secondary focal position, indicating reduced bifocal selectivity. These results demonstrate that mixed-phase conditions lead to incomplete phase modulation and, consequently, degraded focal contrast. Therefore, precise control of the GST phase transition is essential for achieving optimal bifocal performance.}
\newtext{The axial intensity maps in  \textbf{Fig.~\ref{fig:intensity}} and\textbf{~\ref{fig:pol}} were first obtained using 2D FDTD simulations to efficiently visualize the full-lens focusing evolution in the two GST states. To verify that the same focusing behavior is preserved in the actual three-dimensional device, additional full-lens 3D FDTD simulations were performed. The corresponding intensity profiles in \textbf{Fig.~\ref{fig:row_images}} confirm the focusing behavior and show good agreement with the 2D results, supporting the validity of the adopted approach.}

\text{Table~\ref{tab:pcm_varifocal}} compares recently reported varifocal metalenses based on different phase-change materials. 
While Sb$_2$Se$_3$, GSST, and plasmonic Au--GST designs exhibit limited efficiency or polarization dependence, 
previous GST-based metasurfaces also suffered from optical losses and narrow tunability. 
In contrast, the present all-dielectric GST design achieves dual focal lengths (70~$\mu$m and 200~$\mu$m) 
with higher efficiency (20--30\%) and polarization insensitivity, highlighting its superiority in low-loss, 
fast reconfigurable photonic performance. \newtext{It should be noted that the comparison of performance metrics across different tunable metasurface designs is influenced by several factors. The proposed hybrid Si--GST structure differs from pure-GST and plasmonic metasurfaces in terms of optical confinement, material absorption, and modal distribution, all of which affect focusing efficiency and switching behavior. Furthermore, the switching times reported in this work are based on numerical thermal simulations, whereas many previously reported values were obtained experimentally under different excitation and thermal processing conditions. In addition, focusing efficiency is not defined identically across studies and may vary depending on the choice of focal-region area and the normalization with respect to transmitted or incident power. Therefore, the comparison presented in \text{Table~\ref{tab:pcm_varifocal}} is intended to provide qualitative insight into general performance trends rather than direct quantitative equivalence.}

\begin{table*}[t]
\centering
\caption{Comparison of tunable and varifocal metalenses based on phase-change materials.}
\label{tab:pcm_varifocal}
\small
\renewcommand{\arraystretch}{1.3}
\begin{tabularx}{\textwidth}{@{}l l c c c c@{}}
\toprule
\textbf{Platform} & \textbf{Tuning Mechanism} & $\boldsymbol{\lambda}$ (\textmu m) & \textbf{Pol.} & \textbf{Eff. (\%)} & \textbf{Switching Time} \\
\textbf{(Exp.)/(Num.)*} & & & & & \\
\midrule

\makecell[l]{\textcolor{red}{(Exp.)} GSST\\on CaF$_2$\cite{shalaginov2021reconfigurable}} & \makecell[l]{Hot-plate\\annealing} & 5.2 & LP & 21--23 & 30 min \\
\midrule

\makecell[l]{\textcolor{red}{(Exp.)} Sb$_2$Se$_3$\\metasurface\cite{wang2023varifocal}} & \makecell[l]{Optical\\(laser heating)} & 1.55 & LP/CP & 3--5.7 & 5 min \\
\midrule

\makecell[l]{\textcolor{red}{(Exp.)} GST on\\plasmonic\\metasurfaces\cite{yin2017beam}} & \makecell[l]{Electro-\\thermal} & 1.55 & CP & 5--10 & 2 min \\
\midrule

\makecell[l]{\textcolor{red}{(Exp.)} Sb$_2$Se$_3$\\on ITO\cite{shen2023enhancing}} & \makecell[l]{Electro-\\thermal} & 1.55 & LP & -- & 2--8 $\mu$s \\
\midrule

\makecell[l]{\textcolor{red}{(Num.)} GST\\on Si\cite{r17}} & \makecell[l]{Thermal\\tuning} & 1.55 & LP & 7--8 & -- \\
\midrule

\makecell[l]{\textcolor{red}{\textbf{(Num.)}} ]\textbf{Hybrid Si--GST}\\\textbf{(this work)}} & \makecell[l]{\textbf{All-optical}\\\textbf{(flat-top laser)}} & \textbf{1.55} & \textbf{LP/CP} & \textbf{20--30} & \textbf{8--90 ns} \\
\bottomrule
\end{tabularx}

\vspace{0.2cm}
\raggedright {\footnotesize \textcolor{red}{*Exp. -- Experimental Study, Num. -- Numerical Study}}

\end{table*}

\subsection{Laser-Induced Phase Transition in GST-Based Meta-Atoms}

Optical control of transmittance modulation in the proposed metasurface was achieved through laser-induced phase transitions in the Ge$_2$Sb$_2$Te$_5$ (GST) layer embedded within two distinct meta-atom architectures, illustrated in \textbf{Fig.~\ref{fig:img1}}. An accurate description of the laser-induced heat source is crucial for modeling thermal behavior during processing. Although the \textit{fundamental Gaussian beam} is widely used to describe laser intensity distributions~\cite{kundakcioglu2016transient}, it provides an idealized representation that may not be valid for all practical cases, particularly when the beam quality is moderate~\cite{yoshida1996propagation}. To address this, several alternative models have been introduced in the literature, including the \textit{double-ellipsoid power density distribution} proposed by Goldak~\cite{goldak1984new} and \textit{flat-top beam} formulations~\cite{paschotta2008encyclopedia}.

In this study, a \textit{super-Gaussian profile}, a smoothed flat-top distribution was employed to represent the transverse intensity of the laser beam. The heat generation due to this super-Gaussian beam of order $n$ is expressed as~\cite{x1}:
\begin{equation}
Q(r) = Q_0 \exp \left[-2 \left( \frac{r}{w_0} \right)^n \right]
\label{eq:supergaussian}
\end{equation}
where $Q_0$ denotes the peak intensity, $w_0$ the beam radius at the surface, and $r$ the radial distance from the beam axis. A standard Gaussian beam corresponds to $n = 2$, whereas higher orders produce flatter intensity profiles with sharper boundaries. Under these conditions, and for a laser operating at power $P$, the peak intensity becomes:
\begin{equation}
Q_0 = \frac{P}{\pi w_0^2}
\label{eq:peakintensity}
\end{equation}

 A flat-top laser beam was used to provide spatially uniform illumination, ensuring even optical absorption across the nanofin surface. This flat-top intensity distribution minimizes localized overheating, leading to homogeneous temperature evolution and reliable switching between amorphous and crystalline states~\cite{du2021raman}.

During the crystallization process, the GST layer is heated above its glass transition temperature ($T_\mathrm{g} \approx 650$~K) using a moderate-power, long-duration flat-top laser pulse (hundreds of nanoseconds to microseconds)~\cite{orava2012characterization}. In Meta-atom~1, heat is primarily absorbed within the GST and subsequently diffuses through the Al$_2$O$_3$ substrate, which stabilizes the thermal profile and prevents excessive gradients. In contrast, Meta-atom~2 benefits from the inclusion of an intermediate Si pillar, which enhances vertical heat transport and facilitates faster, more uniform crystallization with reduced mechanical stress within the GST region. For amorphization, the GST must be rapidly cycled above its melting temperature ($T_\mathrm{m} \approx 900$~K), followed by a fast quenching process. This is achieved using a high-intensity, short-duration flat-top laser pulse. The GST layer melts uniformly under irradiation and then cools rapidly back into the amorphous phase~\cite{x2}.

\begin{figure}[htbp]
    \centering

    \begin{subfigure}[b]{0.37\linewidth}
        \centering
        \includegraphics[width=\linewidth]{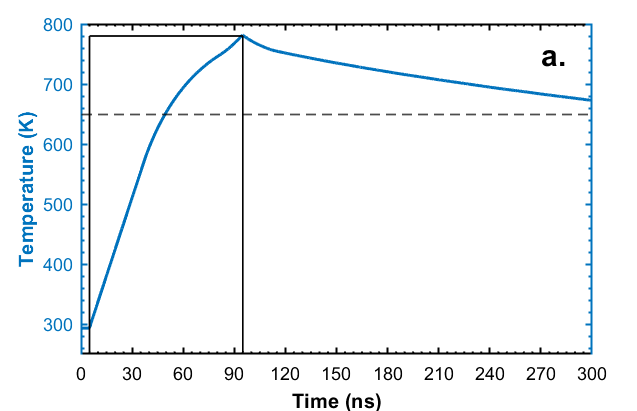}
        
        \label{fig:img11}
    \end{subfigure}
    \hspace{2mm}
    \begin{subfigure}[b]{0.37\linewidth}
        \centering
        \includegraphics[width=\linewidth]{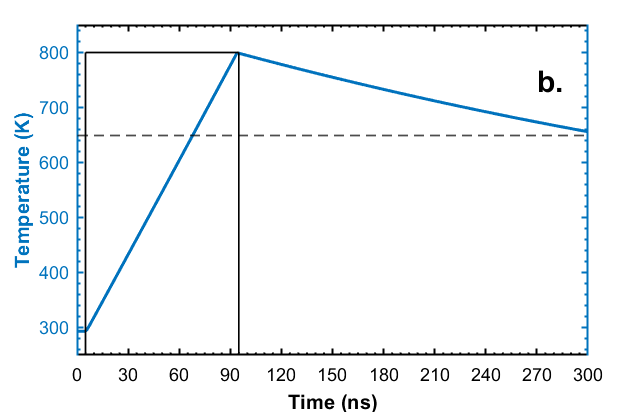}
     
        \label{fig:img12}
    \end{subfigure}

    \vspace{3mm} 

    \begin{subfigure}[b]{0.37\linewidth}
        \centering
        \includegraphics[width=\linewidth]{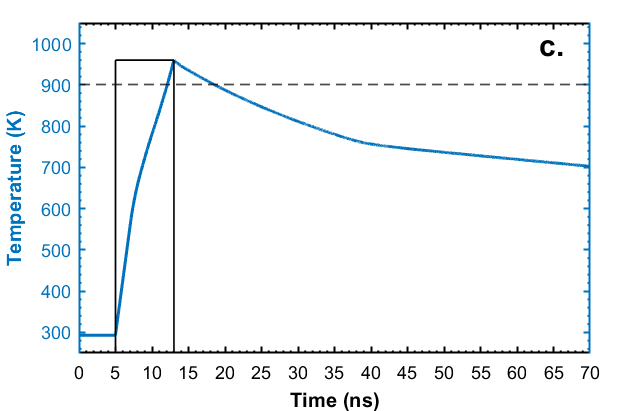}

        \label{fig:img7}
    \end{subfigure}
    \hspace{2mm}
    \begin{subfigure}[b]{0.37\linewidth}
        \centering
        \includegraphics[width=\linewidth]{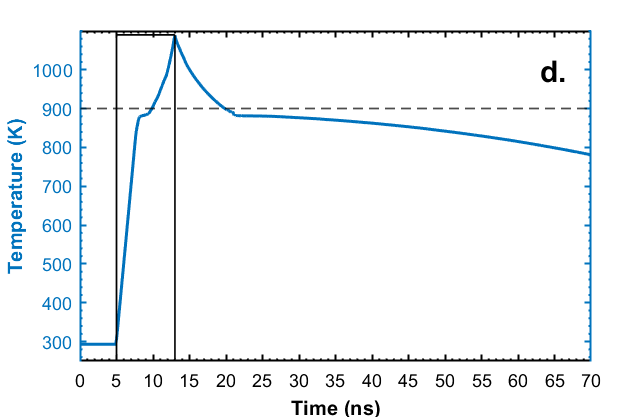}

        \label{fig:img10}
    \end{subfigure}

    \caption{Simulated average temperature evolution of nanopillars under laser excitation. 
(a) and (c) correspond to GST-only nanopillars, while (b) and (d) correspond to hybrid GST-Si nanopillars. 
In (a) and (b), a 10~mW, 90~ns flat-top laser pulse gradually heats the GST layer toward its glass-transition temperature ($T_\mathrm{g} \approx 650~\mathrm{K}$), initiating crystallization through atomic rearrangement into the ordered phase. 
In (c) and (d), a 90~mW, 8~ns laser pulse briefly elevates the temperature above the melting point ($T_\mathrm{m} \approx 900~\mathrm{K}$), followed by rapid quenching that restores the amorphous state. 
The horizontal dashed lines in (a) and (b) mark the moment when the temperature exceeds $T_\mathrm{g}$, signifying the onset of crystallization, whereas in (c) and (d) the transient temperature surpasses $T_\mathrm{m}$, initiating melting. }
    \label{fig:heat}
\end{figure}

\begin{figure*}[htbp]
  \centering

  \begin{subfigure}[b]{0.23\textwidth}
    \centering
    \includegraphics[width=\linewidth]{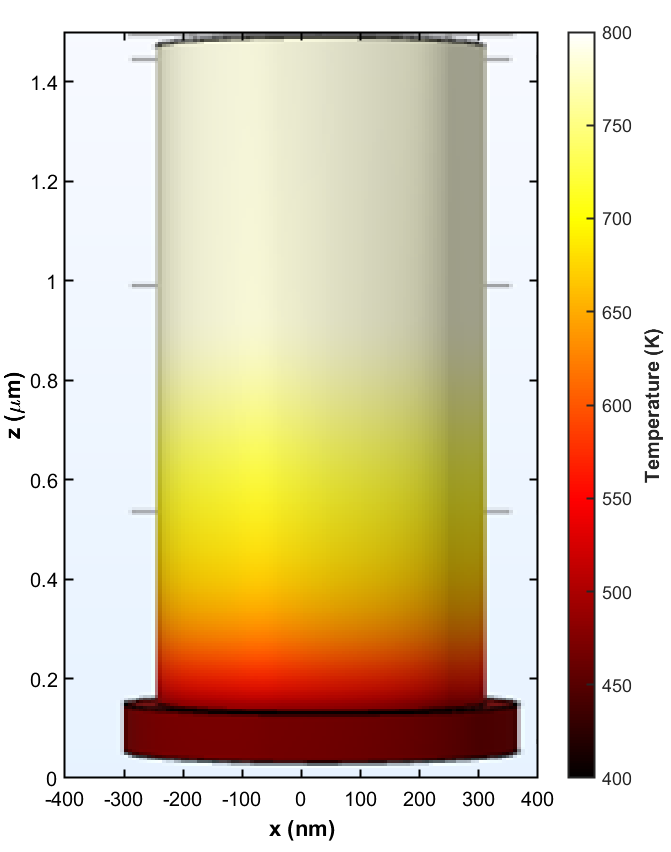}
    \caption{}
    \label{fig:img19}
  \end{subfigure}\hfill
  \begin{subfigure}[b]{0.23\textwidth}
    \centering
    \includegraphics[width=\linewidth]{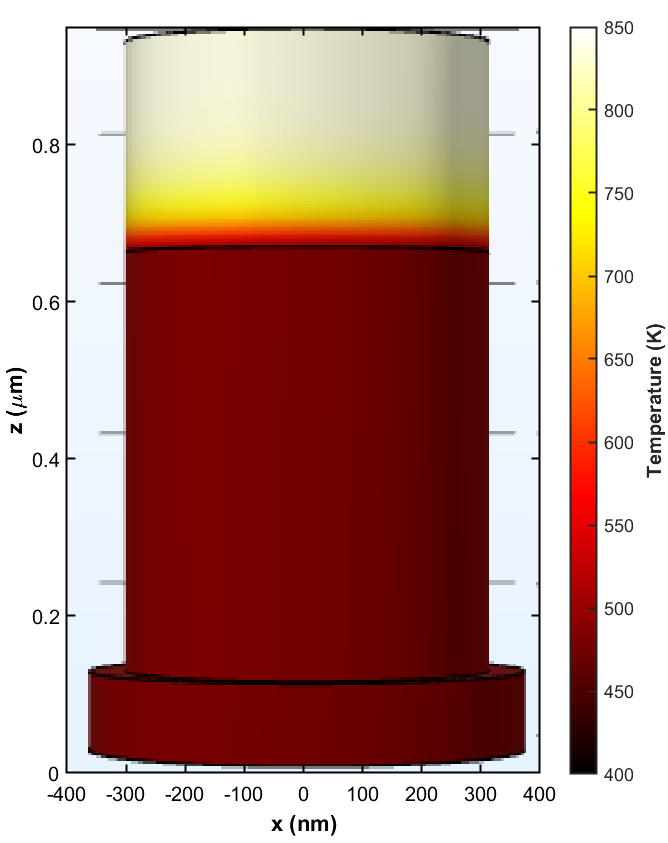}
    \caption{}
    \label{fig:img20}
  \end{subfigure}\hfill
  \begin{subfigure}[b]{0.23\textwidth}
    \centering
    \includegraphics[width=\linewidth]{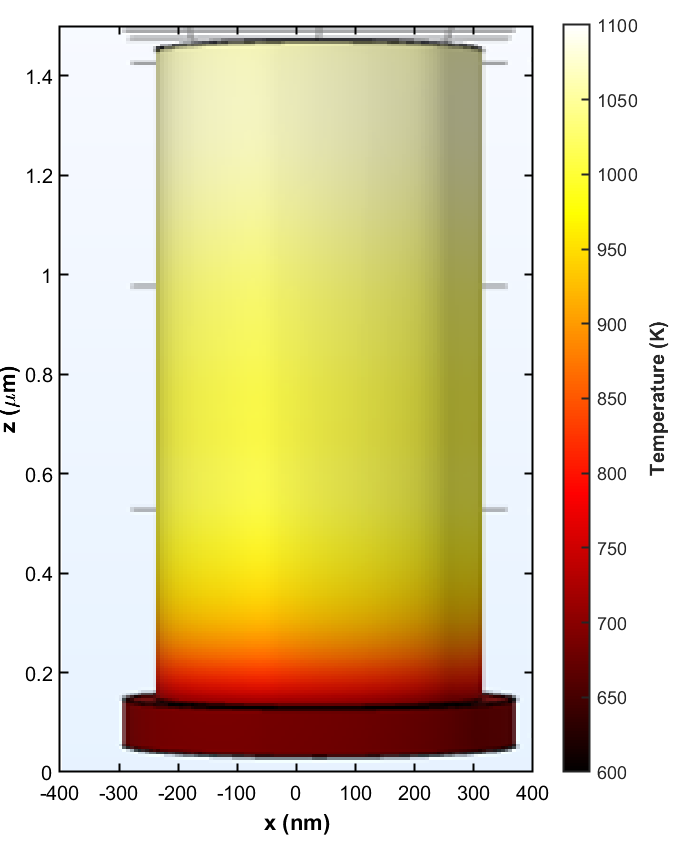}
    \caption{}
    \label{fig:img17}
  \end{subfigure}\hfill
  \begin{subfigure}[b]{0.23\textwidth}
    \centering
    \includegraphics[width=\linewidth]{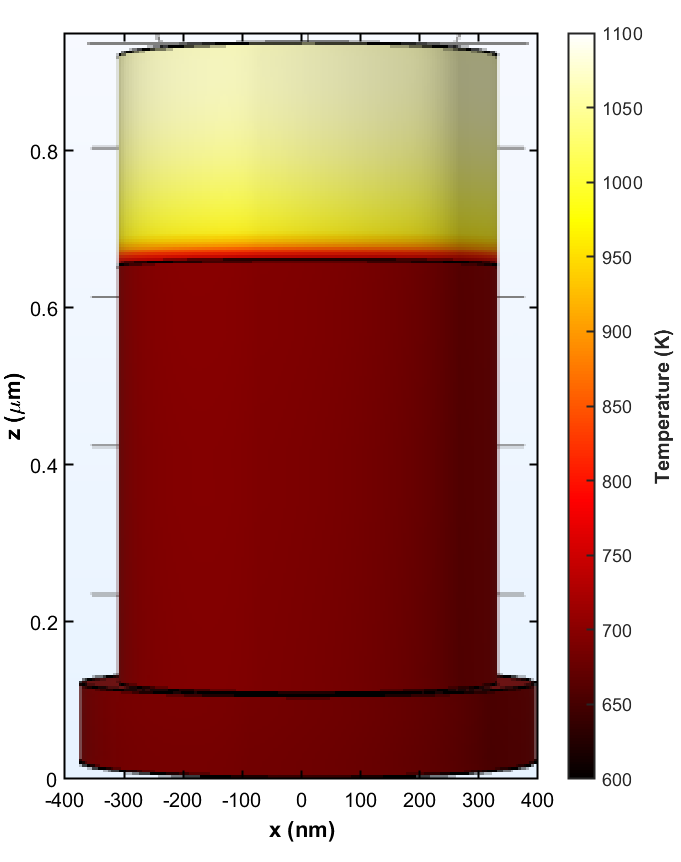}
    \caption{}
    \label{fig:img18}
  \end{subfigure}

  \caption{Simulated temperature distribution in GST nanopillars during laser excitation. 
(a, c) correspond to GST-only and (b, d) to hybrid GST-Si nanopillar configurations under crystallization and amorphization processes, respectively, showing uniform and stable thermal evolution across the structures. }
  \label{fig:temp}
\end{figure*}

The Radiative Beam in Absorbing Media and Heat Transfer in Solids modules were coupled in COMSOL Multiphysics to perform numerical simulations. All external surfaces of the structure were assigned surface-to-ambient radiation boundary conditions along with a convective heat flux to account for natural air convection, assuming a heat transfer coefficient of $h = 10~\mathrm{W\,m^{-2}K^{-1}}$~\cite{heigel2015thermo}. The initial temperature of the model was set to $273.15~\mathrm{K}$. The thermal properties of GST and Si used in the simulations are provided in the \href{Sup_Doc.pdf}{Supplement 1}. The GST nanofin has been exposed to a flat-top laser beam ($n \approx 20$
) at 1030 nm with a uniform intensity profile and a spot size of 300 nm~\cite{caiazzo2018simulation}. During crystallization, the GST with a radius of 250~nm nanopillars of both types are gradually heated to near its crystallization temperature ($T_\mathrm{g}$) using a 10~mW, 90~ns laser pulse, as illustrated in \textbf{Fig.~\ref{fig:heat}}a and \textbf{Fig.~\ref{fig:heat}}b. For amorphization, a 90~mW, 8~ns pulse momentarily elevates the temperature above the melting point ($T_\mathrm{m}$) before rapid cooling restores the amorphous phase, as shown in \textbf{Fig.~\ref{fig:heat}}c and \textbf{Fig.~\ref{fig:heat}}d. These simulations were performed for both nanopillar architectures with a diameter of 250~nm, and the corresponding temperature distributions for the crystallization and amorphization processes are presented in \textbf{Fig.~\ref{fig:temp}}a, \textbf{Fig.~\ref{fig:temp}}b and \textbf{Fig.~\ref{fig:temp}}c, \textbf{Fig.~\ref{fig:temp}}d, respectively.  The estimated optical energy required to induce phase transitions is approximately $3.53~\mathrm{nJ/\mu m^{2}}$ for crystallization and $4.13~\mathrm{nJ/\mu m^{2}}$ for amorphization, demonstrating the energy-efficient operation of the proposed metasurface.\color{red} \sout{The nearly uniform temperature distribution across both nanopillar architectures confirms the stability and spatial homogeneity of the phase transition process. However, the crystallization of GST is relatively slower, serving as the primary speed-limiting factor of the device}~\cite{senkader2004models}.\color{red} \sout{ Consequently, the overall performance of the phase-change device is governed by the crystallization kinetics of the GST layer. As observed in \textbf{Fig.~\ref{fig:heat}},} \color{red} \sout{the amorphous GST requires approximately 90~ns to achieve complete crystallization, implying a maximum theoretical switching frequency of \textasciitilde 10~MHz.
The temperature stays constant throughout the nanofin due to the flat-top illumination, guaranteeing steady phase transitions free from thermal damage.} \newtext{The crystallization process of GST is intrinsically slower than amorphization due to its thermally activated nucleation-and-growth mechanism. In the present design, the simulated crystallization time is approximately $90$~ns, corresponding to an estimated maximum switching frequency on the order of $10$~MHz. This value should be interpreted as a simulation-based estimate for the proposed metasurface geometry and thermal conditions, rather than a universal theoretical limit of GST. In addition, the structural configuration of the metasurface can influence the switching dynamics: although the Si layer improves optical performance and transmission efficiency, it may also modify the thermal dissipation pathway and thereby affect the crystallization speed. A more rigorous kinetic analysis, including activation-energy extraction and experimental time-resolved validation, is beyond the scope of the present work and remains an important direction for future study.} \newtext{The thermal simulations are based on a simplified one-way multiphysics model, in which the absorbed optical power is treated as the heat source. Bidirectional electromagnetic--thermal coupling, temperature-dependent refractive-index changes, and dynamic thermal feedback during pulsed operation are not included. Since the proposed metasurface is primarily all-dielectric, Joule heating from metallic components is expected to be negligible. Although this approach captures the dominant heating mechanism and provides reasonable estimates of the switching time, a fully coupled model would offer a more comprehensive description and is reserved for future work. The long-term stability and practical applicability of the proposed metasurface are influenced by factors such as thermal cycling endurance, environmental conditions, and device-level thermal management. While the present study focuses on simulation-based analysis, GST-based phase-change materials have demonstrated high cycling stability and repeatable switching behavior in previous experimental works. The performance of the metasurface under repeated switching cycles and varying environmental conditions (e.g., temperature fluctuations and humidity) may affect the optical response and efficiency retention. In addition, alternative operation schemes, such as self-powered or light--thermal feedback mechanisms, could further enhance the practicality of the device. These aspects will be explored in future experimental studies.}

 \color{black}
\section{Conclusion}

In conclusion, our work proposes a phase-change chalcogenide metasurface-based dynamically reconfigurable varifocal metalens. The two concentric zones in the design—the outer GST-only zone and the inner hybrid Si–GST zone are each designed to generate distinct optical phase responses that correlate to short ($f_{1}=70~\mu$m)  and long ($f_{2}=200~\mu$m) focal lengths. The device achieves full $0$–$2\pi$ phase coverage with great transmission efficiency by optimizing the nanopillar structure using FDTD. With focusing efficiencies of about 20\% and 30\% for $f_{1}$ and $f_{2}$, respectively, the metalens effectively alternates between the two focal states when the GST changes between its crystalline and amorphous phases.
We demonstrate that flat-top laser illumination enables consistent and reversible phase transitions in the GST layer, offering precise optical control without any requirement for mechanical movement, is further confirmed by coupled electromagnetic–thermal simulations. 
Overall, this study offers a viable path toward the development of compact, nonvolatile, high-speed reconfigurable metalenses. The suggested method provides a versatile platform for adaptive imaging, LiDAR, and optical communication,  and integrated photonic systems that increasingly require dynamic beam shaping and variable focusing capabilities by fusing hybrid dielectric PCM meta-atoms with spatially controlled laser excitation. In conclusion, We have presented a phase-change chalcogenide metasurface metalens capable of rapid, all-optical varifocal switching through spatially controlled laser heating. By integrating GST-Si hybrid nanofins with flat-top optical excitation, the device achieves nonvolatile, reversible switching between dual focal states while maintaining polarization-insensitive and near-diffraction-limited performance. The combined optical–thermal simulations confirm stable phase transitions, uniform temperature distribution, and sub-microsecond operation with low energy consumption. This approach establishes a scalable platform for reconfigurable flat-optics, merging high-index dielectric design with optical phase-change control. Future work will focus on experimental fabrication and characterization to realize high-speed tunable imaging, beam steering, and photonic signal routing within integrated optical systems.








\begin{backmatter}
\bmsection{Funding}
This research received no specific grant.
\bmsection{Acknowledgments}

\bmsection{Disclosures}
The authors declare no conflict of interest.

\bmsection{Data availability} Data underlying this work are not publicly available at this time but will be provided upon reasonable request.

\bmsection{Supplemental document}

\section*{MATERIAL DATA}

The wavelength-dependent complex refractive indices used in the simulations are shown in \textbf{Fig.~\ref{fig:imgs1}}. The real and imaginary components were obtained from experimentally validated datasets for Ge$_2$Sb$_2$Te$_5$ (GST) \cite{rs1}. Additional properties of the materials (GST, Si, and Al$_2$O$_3$), including density, thermal conductivity, and specific heat capacity, are summarized in Table~\ref{tab:properties}.

\begin{figure}[htbp]
    \centering
    \includegraphics[width=0.7\linewidth]{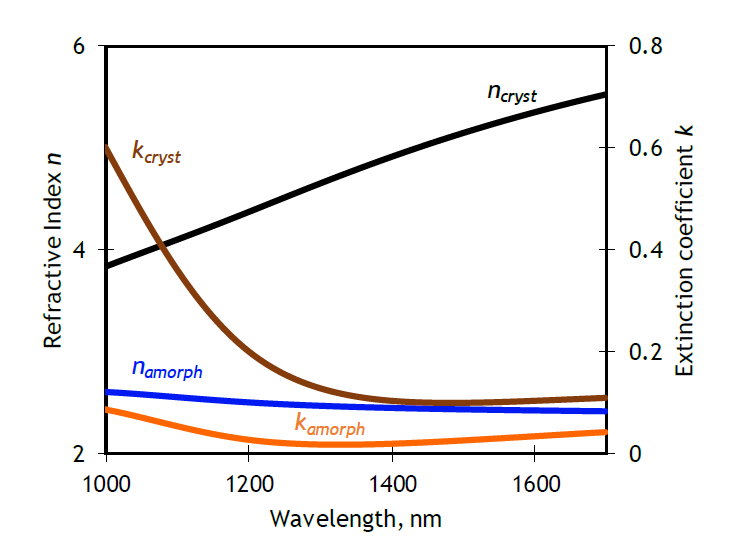}
    \caption{Wavelength-dependent refractive index ($n$) and extinction coefficient ($k$) of amorphous and crystalline GST \cite{rs1}.}
    \label{fig:imgs1}
\end{figure}

\begin{table}[h!]
\centering
\caption{Material parameters for the device simulation.}
\label{tab:properties}
\begin{tabular}{lccc}
\hline
\textbf{Material} & \textbf{Density (kg·m$^{-3}$)} & \textbf{Specific Heat (J·K$^{-1}$·kg$^{-1}$)} & \textbf{Thermal Conductivity (W·m$^{-1}$·K$^{-1}$)} \\
\hline
Si     & 2330\cite{udrea2001design} & 700\cite{udrea2001design} & 157\cite{udrea2001design} \\
Al$_2$O$_3$ & 3970\cite{devi2017heat} & 765\cite{devi2017heat} & 40\cite{devi2017heat} \\
GST (amorphous)  & 6150\cite{wright2010design} & 210\cite{carrillo2016design} & 0.20\cite{carrillo2016design} \\
GST (crystalline) & 6150\cite{wright2010design} & 210\cite{carrillo2016design} & 0.58\cite{carrillo2016design} \\
\hline
\end{tabular}
\end{table}

\section*{DESIGN OPTIMIZATION OF GST META-ATOMS}

\textbf{Fig.~\ref{fig:S9}}a, \textbf{Fig.~\ref{fig:S9}}b, and \textbf{Fig.~\ref{fig:S9}}c illustrate the structure and optical performance of the all-GST nanopillar meta-atom.
The design consists of a single GST pillar of height $H_{1}$, diameter $D$, and lattice period $P$ fabricated on a sapphire substrate.
The transmittance and phase maps demonstrate the dependence of the optical response on the GST radius and height.
The optimal GST height, highlighted by the black dashed line, was selected to achieve high transmittance and complete $2\pi$ phase modulation, ensuring efficient phase control in the amorphous state.

\section*{DESIGN OPTIMIZATION OF HYBRID GST--SI META-ATOMS}

The structural parameters of the proposed hybrid GST--Si nanopillar unit cell are illustrated in \textbf{Fig.~\ref{fig:S9}}d.
The pillar consists of a GST segment of height $H_{2}$ stacked over a Si base of height $H_{3}$, both resting on a sapphire substrate with lattice period $P$ and pillar diameter $D$.
To determine the optimal design parameters that provide high transmittance and complete $2\pi$ phase coverage, parametric simulations were performed by varying the GST radius and height.
\textbf{Fig.~\ref{fig:S9}}e and \textbf{Fig.~\ref{fig:S9}}f show the corresponding transmittance and phase responses, respectively.
As indicated by the black dashed line, the selected GST height $H_{2}$ represents the value adopted throughout this work, offering an effective trade-off between transmission efficiency and continuous phase modulation.

\begin{figure}[htbp]
    \centering
    \begin{subfigure}[b]{0.17\linewidth}
        \centering
        \includegraphics[width=\linewidth]{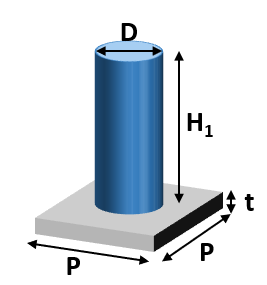}
        \caption{}
        \label{fig:img_a}
    \end{subfigure}
    \hfill
    \begin{subfigure}[b]{0.35\linewidth}
        \centering
        \includegraphics[width=\linewidth]{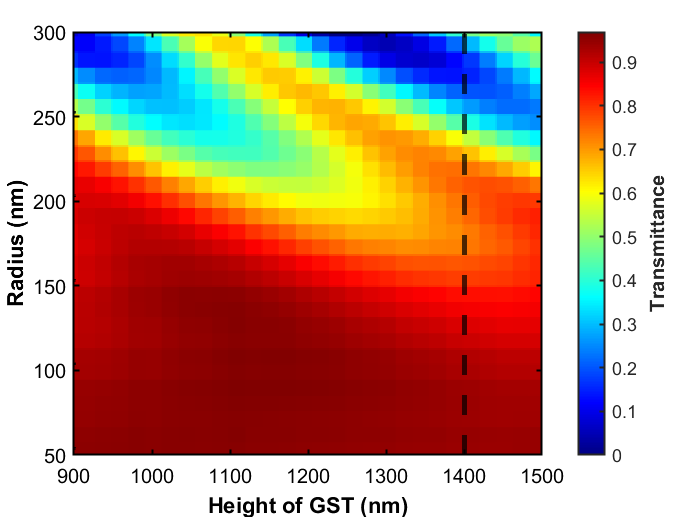}
        \caption{}
        \label{fig:img_b}
    \end{subfigure}
    \hfill
    \begin{subfigure}[b]{0.35\linewidth}
        \centering
        \includegraphics[width=\linewidth]{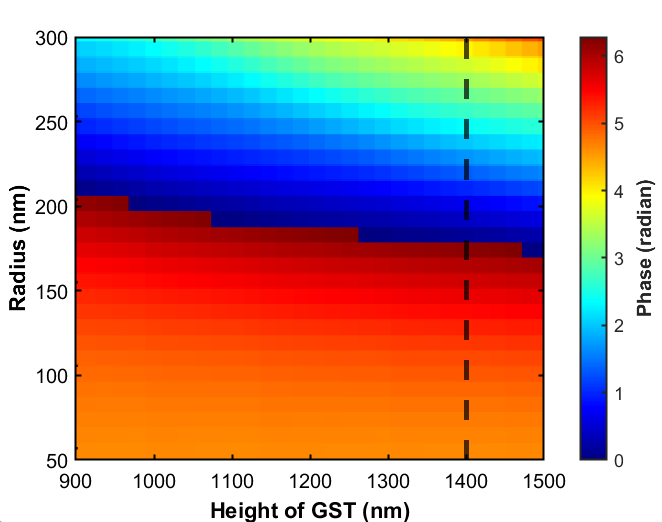}
        \caption{}
        \label{fig:img_c}
    \end{subfigure}

    \vspace{5pt}
    \begin{subfigure}[b]{0.17\linewidth}
        \centering
        \includegraphics[width=\linewidth]{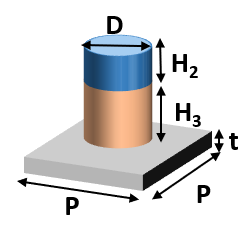}
        \caption{}
        \label{fig:img_d}
    \end{subfigure}
    \hfill
    \begin{subfigure}[b]{0.35\linewidth}
        \centering
        \includegraphics[width=\linewidth]{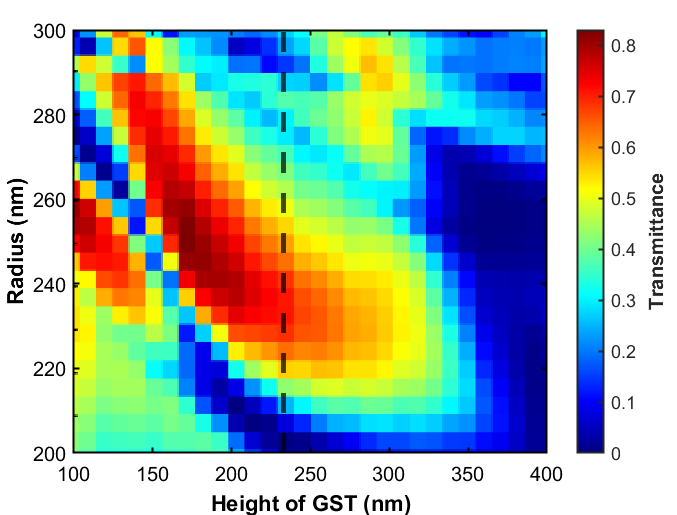}
        \caption{}
        \label{fig:img_e}
    \end{subfigure}
    \hfill
    \begin{subfigure}[b]{0.35\linewidth}
        \centering
        \includegraphics[width=\linewidth]{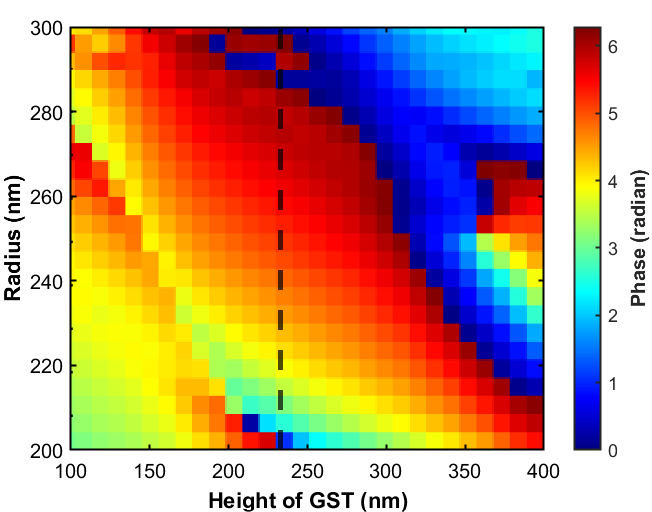}
        \caption{}
        \label{fig:img_f}
    \end{subfigure}

    \caption{(a--c) Simulated schematic, transmittance, and phase response of the all-GST meta-atom as functions of pillar radius and height. (d--f) Corresponding schematic, transmittance, and phase response of the hybrid GST--Si meta-atom. The black dashed lines indicate the optimized GST height selected for achieving full $2\pi$ phase coverage with high transmittance.}
    \label{fig:S9}
\end{figure}

\section*{Phase-graded metasurface}

Tables~\ref{tab:amorphous_phase_d} and~\ref{tab:crystalline_phase_d} present the numerical datasets used to define the nanopillar radii corresponding to discrete optical phase levels for the metasurface design. Each table contains a set of 24 uniformly distributed phase levels ranging from 0$^{\circ}$ to 345$^{\circ}$ in 15$^{\circ}$ increments, together with their associated nanopillar radii ($R$) in nanometers. The resulting quantized phase map reproduces the designed phase progression with high fidelity, ensuring smooth phase continuity across the metasurface aperture while minimizing quantization-induced phase errors.

\begin{table}[ht!]
\centering
\caption{Dataset of the relative phases corresponding to the nanopillar radius (Outer region).}
\label{tab:amorphous_phase_d}
\renewcommand{\arraystretch}{1.1}
\setlength{\tabcolsep}{8pt}
\begin{tabular}{cccc}
\toprule
\textbf{Phase (deg)} & \textbf{R (nm)} & \textbf{Phase (deg)} & \textbf{R (nm)} \\
\midrule
0   & 81.02  & 180 & 217.35 \\
15  & 108.27 & 195 & 225.21 \\
30  & 127.25 & 210 & 233.25 \\
45  & 143.75 & 225 & 240.61 \\
60  & 157.48 & 240 & 247.57 \\
75  & 168.41 & 255 & 256.18 \\
90  & 178.60 & 270 & 264.19 \\
105 & 186.94 & 285 & 271.65 \\
120 & 195.63 & 300 & 279.41 \\
135 & 204.21 & 315 & 286.36 \\
150 & 211.18 & 330 & 292.48 \\
165 & 217.35 & 345 & 296.64 \\
\bottomrule
\end{tabular}
\end{table}

\begin{table}[ht!]
\centering
\caption{Dataset of the relative phases corresponding to the nanopillar radius (Inner region).}
\label{tab:crystalline_phase_d}
\renewcommand{\arraystretch}{1.1}
\setlength{\tabcolsep}{8pt}
\begin{tabular}{cccc}
\toprule
\textbf{Phase (deg)} & \textbf{R (nm)} & \textbf{Phase (deg)} & \textbf{R (nm)} \\
\midrule
0   & 200.65 & 180 & 221.88 \\
15  & 201.32 & 195 & 227.17 \\
30  & 202.04 & 210 & 232.79 \\
45  & 202.99 & 225 & 239.90 \\
60  & 204.08 & 240 & 246.89 \\
75  & 205.36 & 255 & 255.57 \\
90  & 206.82 & 270 & 264.68 \\
105 & 208.44 & 285 & 276.05 \\
120 & 211.18 & 300 & 281.72 \\
135 & 214.04 & 315 & 292.70 \\
150 & 217.96 & 330 & 296.93 \\
165 & 221.88 & 345 & 299.84 \\
\bottomrule
\end{tabular}
\end{table}

\section*{Phase Distribution Analysis of the Varifocal Metalens}

As shown in \textbf{Fig.~\ref{fig:target}}a, a comparison is presented between the simulated and theoretical phase profiles of the designed varifocal metalens under the two phase states of GST. In the amorphous state, the outer GST region (Region~2, 35--55~$\mu$m) satisfies the phase-matching condition corresponding to a longer focal length of $f_{2} = 200~\mu$m, while the inner hybrid GST--Si region remains optically inactive. Conversely, when GST transitions to the crystalline state, as depicted in \textbf{Fig.~\ref{fig:target}}b, the inner hybrid region (Region~1, radius~$\leq$~35~$\mu$m) provides the required phase modulation to realize a shorter focal length of $f_{1} = 70~\mu$m, whereas the outer GST region becomes inactive. The blue curves in both figures represent the simulated phase response, while the green curves correspond to the theoretical hyperboloidal target phase distribution.

\begin{figure}[htbp]
    \centering
    \begin{subfigure}[b]{0.48\linewidth}
        \centering
        \textbf{(a)}\\[0pt]
        \includegraphics[width=\linewidth]{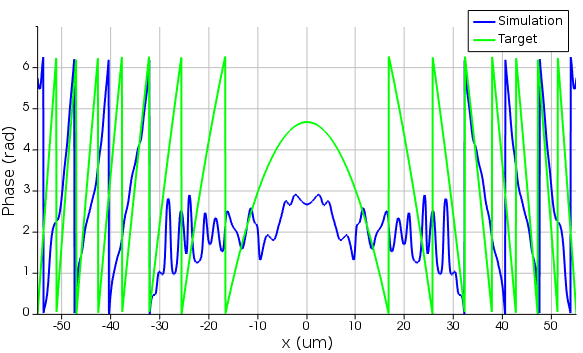}
    \end{subfigure}
    \hfill
    \begin{subfigure}[b]{0.48\linewidth}
        \centering
        \textbf{(b)}\\[0pt]
        \includegraphics[width=\linewidth]{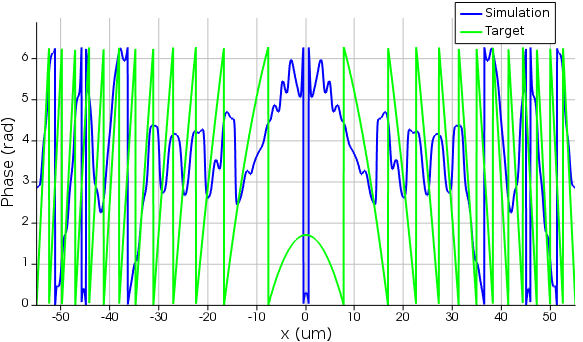}
    \end{subfigure}
    \caption{Comparison between the simulated and target phase distributions of the designed varifocal metalens for the two GST states: (a) amorphous phase and (b) crystalline phase.}
    \label{fig:target}
\end{figure}

\section*{THERMAL SIMULATION OF LASER-INDUCED GST}

The thermal response of GST during laser irradiation was numerically simulated using \textit{COMSOL Multiphysics} \cite{darif2008numerical}. This simulation aimed to demonstrate the phase change from amorphous to crystalline and vice versa. The transient heat transfer within the solid medium is governed by the heat diffusion equation:
\begin{equation}
\rho C_p \frac{\partial T(x,y,z,t)}{\partial t} = \kappa \nabla T(x,y,z,t) + Q(x,y,z,t),
\tag{E1}
\end{equation}
where $\rho$ is the material density, $C_p$ is the specific heat capacity, $\kappa$ is the thermal conductivity, $T$ is the temperature, and $Q(x,y,z,t)$ represents the heat source of the top-hat laser beam. The absorbed laser energy follows the Beer--Lambert law and can be expressed as:
\begin{equation}
\label{eq:E2}
Q(x,y,z,t) = (1 - R)\,\alpha\,e^{-\alpha z}\,\frac{P}{l^2}\,f(t),
\tag{E2}
\end{equation}
where $R$ is the surface reflectivity of GST, $\alpha$ is the absorption coefficient, $P$ is the incident laser power, $l$ is the side length of the laser spot, and $f(t)$ represents the temporal pulse profile. The laser beam was assumed to propagate along the $z$-axis.

During amorphization, the \textit{c}-GST layer undergoes melting and rapid solidification into the amorphous phase (\textit{a}-GST). Since the latent heat of fusion significantly affects the temperature evolution during this transition, the phase-change process was modeled using the \textit{Phase Change Material} module in COMSOL \cite{liang2023enhanced}, which is based on the following relations \cite{kiselev2021transmissivity}:

\begin{equation}
\rho_{\text{liquid}} = \rho_{\text{solid}},
\tag{E3}
\end{equation}

\begin{equation}
C_p = \theta_1 C_{p,1} + \theta_2 C_{p,2} + L_{1 \rightarrow 2} \frac{\partial \alpha_m}{\partial T},
\tag{E4}
\end{equation}

\begin{equation}
\alpha_m = \frac{1}{2}\frac{\theta_2 - \theta_1}{\theta_2 + \theta_1},
\tag{E5}
\end{equation}

\begin{equation}
\kappa = \theta_1 \kappa_1 + \theta_2 \kappa_2,
\tag{E6}
\end{equation}

\begin{equation}
\theta_1 + \theta_2 = 1,
\tag{E7}
\end{equation}
where $\theta_1$ and $\theta_2$ are the solid and liquid phase fractions, respectively; $C_{p,1}$ and $C_{p,2}$ are the specific heat capacities of the solid and liquid phases; $\kappa_1$ and $\kappa_2$ are the corresponding thermal conductivities; and $L_{1 \rightarrow 2}$ is the latent heat of fusion. The density of the two phases is assumed to remain constant during melting.

Since part of the laser energy is reflected at the GST surface, the absorbed energy was corrected using Fresnel reflection:
\begin{equation}
\label{eq:Ref}
R = \frac{(n - 1)^2 + k^2}{(n + 1)^2 + k^2},
\tag{E8}
\end{equation}
where $n$ and $k$ are the real and imaginary parts of the refractive index of GST, respectively. The wavelength-dependent values of $n$ and $k$ are taken from the experimentally validated datasets in \textbf{Fig.~\ref{fig:imgs1}}. At a laser wavelength of 1030~nm, the reflectivity ($R$) values calculated from Eq.~\ref{eq:Ref} are 0.195 for amorphous GST and 0.359 for crystalline GST.

The absorption coefficient ($\alpha$) is determined from the extinction coefficient ($k$) using:
\begin{equation}
\alpha = \frac{4\pi k}{\lambda}.
\tag{E9}
\end{equation}

In Eq.~\ref{eq:E2}, the temporal laser waveform $f(t)$ is modeled as a square pulse with a duration of 13~ns for amorphization and 90~ns for crystallization. The latent heat of fusion of GST is taken as 128~kJ/kg \cite{scoggin2018modeling}. The laser power ($P$) applied during amorphization and crystallization processes is 90~mW and 10~mW, respectively. The convective and radiative heat losses, $q_c$ and $q_r$, are expressed as follows:

\begin{equation}
q_c = h(T_\infty - T)
\tag{E10}
\end{equation}

\begin{equation}
q_r = \varepsilon \sigma (T_{\text{room}}^4 - T^4)
\tag{E11}
\end{equation}

where $h$ is the convective heat transfer coefficient, $\varepsilon$ is the surface emissivity, and $\sigma$ is the Stefan--Boltzmann constant. $T_\infty$ and $T_{\text{room}}$ represent the gas medium and room temperatures, respectively. In the subsequent simulations, these losses are incorporated as boundary conditions depending on the specific domain considered. Both $T_\infty$ and $T_{\text{room}}$ are assumed to be 273.15~K. The convective heat transfer coefficient is taken as $h = 10~\text{W}/(\text{m}^2\!\cdot\!\text{K})$ \cite{heigel2015thermo}. The emissivity ($\varepsilon$) of the GST layer is taken as 0.15 for the amorphous phase and 0.75 for the crystalline phase, respectively \cite{kang2024laser}. The physical parameters used in the simulation, including those of the Al$_2$O$_3$ substrate, Si layer, and GST layer, are summarized in Table~\ref{tab:properties}.

\section*{FABRICATION STEPS}

\textbf{Fig.~\ref{fig:imgs12}} and \textbf{Fig.~\ref{fig:imgs13}} illustrate the possible fabrication processes for the all-GST and hybrid GST--Si metasurface regions, respectively.
For the all-GST region, a uniform GST film was first deposited on a sapphire substrate via RF sputtering, followed by CSAR e-beam lithography to define nanoscale patterns.
A thin Cr layer was then deposited using thermal evaporation to serve as an etch mask.
After the lift-off process, the patterned Cr layer was used during reactive ion etching (RIE) to form GST nanopillars, and finally, the Cr mask was removed using a Cr etchant \cite{wang2023varifocal}.
For the hybrid GST--Si region, the process began with Si sputtering to form the base layer, followed by a similar CSAR lithography and Cr deposition sequence to pattern the Si nanopillars.
After RIE and mask removal, a GST layer was sputtered on top of the patterned Si array.
The same lithography and etching steps were repeated to define the GST nanopillars, forming the vertically stacked GST--Si hybrid structure \cite{soliman2025design}.

\begin{figure}[htbp]
    \centering
    \includegraphics[width=0.7\linewidth]{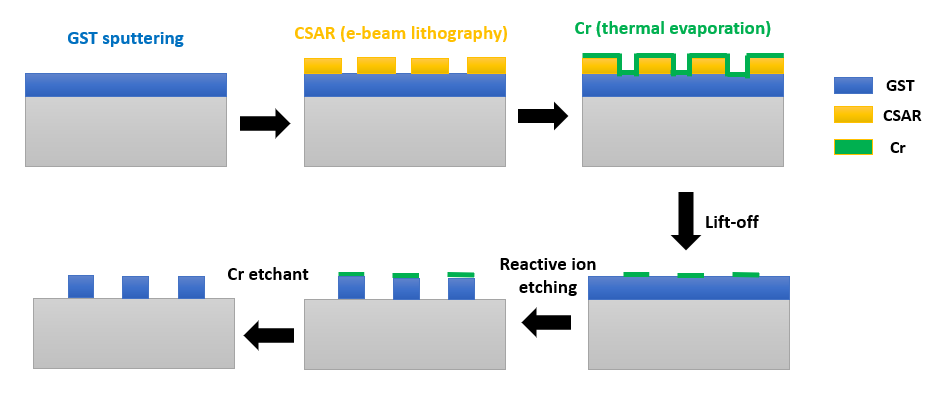}
    \caption{Fabrication process of the all-GST metasurface region.}
    \label{fig:imgs12}
\end{figure}

\begin{figure}[htbp]
    \centering
    \includegraphics[width=0.7\linewidth]{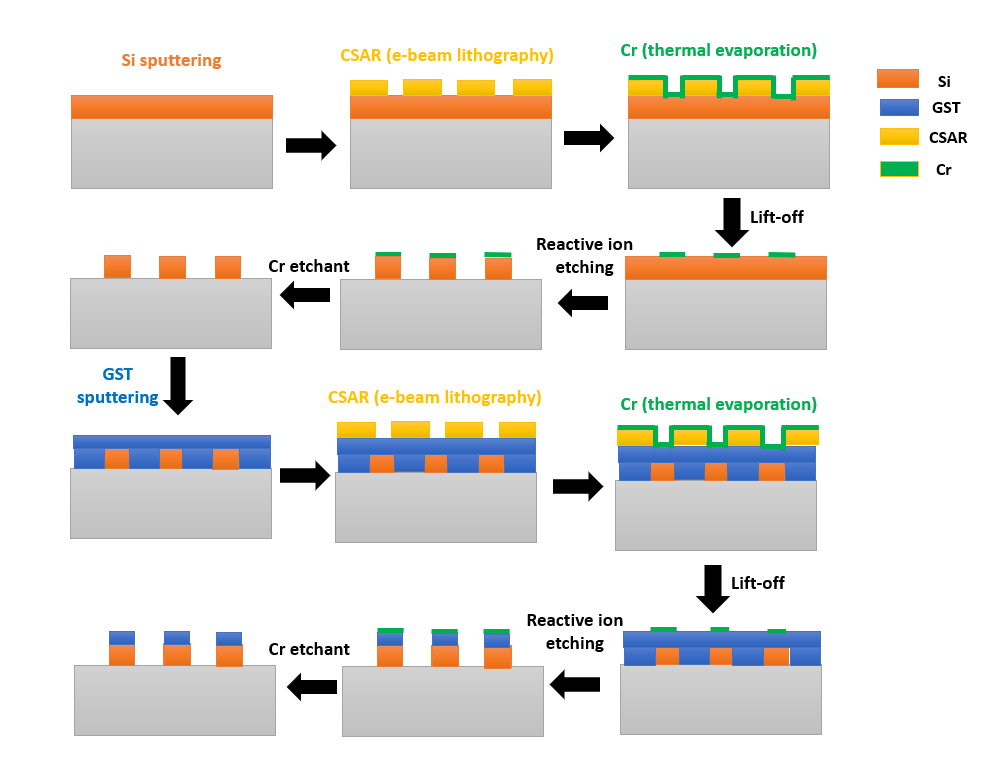}
    \caption{Fabrication process of the hybrid GST--Si metasurface region.}
    \label{fig:imgs13}
\end{figure}
\end{backmatter}
\bibliography{sample}
\end{document}